\documentclass[11pt]{article}
\usepackage{amsthm,amsfonts,amsmath,amssymb,amscd, accents, mathrsfs, mathtools}
\usepackage{authblk}
\usepackage{hyperref}
\usepackage{multicol}
\usepackage{verbatim}
\usepackage{graphicx}

\mathtoolsset{showonlyrefs=false}
                                
\theoremstyle{plain}
\usepackage[numbers]{natbib}
\usepackage[section]{placeins}
\usepackage{float}
\usepackage{program}
\catcode`\|=12\relax
\usepackage{subfigure}
\usepackage[labelsep=space]{caption}
\usepackage{graphicx}% Include figure files
\usepackage{dcolumn}% Align table columns on decimal point
\usepackage{bm}% bold math
\usepackage{etoolbox} % for '\patchcmd' macro
\patchcmd{\subequations}{{0}}{{-1}}{}{}% decrement the equation counter
\patchcmd{\subequations}{\alph}{.\arabic}{}{}
\usepackage{amsmath,amsfonts,amssymb,amsthm}
\usepackage{booktabs} 

\makeatletter
\AtBeginDocument{%
  \expandafter\renewcommand\expandafter\subsection\expandafter{%
    \expandafter\@fb@secFB\subsection
  }%
}
\makeatother

\allowdisplaybreaks

\theoremstyle{thmstyleone}%
\theoremstyle{thmstyletwo}%

\theoremstyle{thmstylethree}%

\author{
Allan Wing-Bocanegra\\
Tecnologico de Monterrey, Escuela de Ingenieria y Ciencias\\
Ave. Eugenio Garza Sada 2501, Monterrey, 64849, N.L., Mexico.\\
  \texttt{A00832476@tec.mx}
  \and
Salvador E. Venegas-Andraca\\
Tecnologico de Monterrey, Escuela de Ingenieria y Ciencias\\
Ave. Eugenio Garza Sada 2501, Monterrey, 64849, N.L., Mexico.\\
  \texttt{svenegas@tec.mx}
}
\title{Circuit Implementation of Discrete-Time Quantum Walks  via the Shunt Decomposition Method}

\begin{document}

\maketitle

%%==================================%%
%% sample for unstructured abstract %%
%%==================================%%

\begin{abstract}
Several models have been proposed to build evolution operators to perform quantum walks in a theoretical way, although when wanting to map the resulting evolution operators into quantum circuits to run them in quantum computers, it is often the case that the mapping process is in fact complicated. Nevertheless, when the adjacency matrix of a graph can be decomposed into a sum of permutation matrices, we can always build a shift operator for a quantum walk that has a block diagonal matrix representation. In this paper, we analyze the mapping process of block diagonal operators into quantum circuit form, and apply this method to obtain quantum circuits that generate quantum walks on the most common topologies found in the literature: the straight line, the cyclic graph, the hypercube and the complete graph. The obtained circuits are then executed on quantum processors of the type Falcon r5.11L and Falcon r4T (two of each type) through IBM Quantum Composer platform and on the Qiskit Aer simulator, performing three steps for each topology. The resulting distributions were compared against analytical distributions, using the statistical distance $\ell_1$ as a performance metric. Regarding experimental executions, we obtained short $\ell_1$ distances in the cases of quantum circuits with a low amount of multi-control gates, being the quantum processors of the type Falcon r4T the ones that provided more accurate results.
\end{abstract}

\maketitle

\section{\label{sec:level1}Introduction}
A random walk is a process describing the dynamics an agent follows by taking random steps on the vertices of a connected graph, i.e. it is a stochastic process formed by successive summation of independent, identically
distributed random variables \cite{lawler_limic_2010}. This mathematical model has been widely used in many areas of science such as in chemistry to study the configurations of polymer-based materials \cite{10.2307/27851819}, in biology to study the movement and dispersal of animals and microorganisms \cite{codling_plank_benhamou_2008}, in computer science to perform image cosegmentation \cite{7299008} and in economics to study the behaviour of stock markets \cite{abakah_alagidede_mensah_ohene-asare_2018}, just to name a few examples and bring to light the versatility of random walks as a mathematical tool. 

Quantum walks constitute a promising universal model of quantum computation originally inspired in random walks, that also studies the dynamics of a walker moving on the vertices of a graph, but in this case the walker can be in a quantum superposition of states and, thus, take multiple directions and have different phases at each step. This brings special properties such as quadratically wider spread of probability distribution \cite{doi:10.1063/1.4903129} and quadratically faster hitting time \cite{krovi_magniez_ozols_roland_2015}, which are key features to obtain a speed-up over classical algorithms. Quantum walks can be either discrete \cite{aharonov2002quantum} or continuous \cite{quantum_decision_trees1998} in time. Within the field of discrete-time quantum walks we can find coined \cite{chandrashekar2010discretetime} or coinless \cite{szegedy_2004, PhysRevA.93.062335} models. In this paper, we focus on the Coined Discrete-Time Quantum Walk model \cite{seva2012}, which is usually just called Discrete-Time Quantum Walk (DTQW) in the literature since it was the first model of its type.

%  still being studied and yet 

The Discrete-Time Quantum Walk model is an active research field that has already proved to be fruitful as it has been used to solve or attempt to solve problems such as image and public-key encryption \cite{yang_pan_sun_xu_2015, public_key_encryption_2015}, the search of elements in an unstructured database \cite{SKB_algoritm, bezerra_lugao_portugal_2021}, the design and training of neural networks \cite{dernbach_mohseni-kabir_pal_gepner_towsley_2019, 8923910}, and  the quantization of the PageRank algorithm \cite{paparo_martin-delgado_2012, pagerank_chawla, PhysRevA.78.012310}, among others, providing a speed-up when compared with random-walk approaches for the same problems. Nevertheless, up to date, quantum walk-based algorithms have mostly been tested theoretically or through simulations \cite{SKB_algoritm, dernbach_mohseni-kabir_pal_gepner_towsley_2019}, and not experimentally due to the fact that quantum computers of the current Noisy Intermediate-Scale Quantum (NISQ) era \cite{preskill_2018}, generate much noise and decoherence, and thus are not suitable to calculate efficiently the probability distributions of quantum walks except for very simple cases, such as few steps of a quantum walk on a line of few nodes as presented by K. N. Cassemiro $et\; al.$ \cite{PhysRevLett.104.050502} and M. A. Broome $et\; al.$ \cite{PhysRevLett.104.153602} on specific quantum processors for this task, and by A. Shakeel \cite{shakeel_2020} and K. Georgopoulos \cite{PhysRevA.103.022408} on general-purpose quantum computers. Efficient experimental implementations of quantum walks on different topologies have been done using other models of quantum walks, i.e. staggered quantum walks \cite{acasiete_agostini_moqadam_portugal_2020}, topological quantum walks \cite{Balu_2018} and continuous-time quantum walks \cite{tang_lin_feng_chen_gao_sun_wang_lai_xu_wang_2018, qiang_loke_montanaro_aungskunsiri_zhou_obrien_wang_matthews_2016, Jiao:21}, being one key factor, according to \cite{acasiete_agostini_moqadam_portugal_2020}, the fact that coinless (e.g. staggered, topological or continuous) quantum walks do not need an extra register for the coin state, and in NISQ computers additional qubits induce more noise to the system, thus reducing fidelity in experimental runs.

As a step forward to efficiently run DTQW-based algorithms on quantum computers, in this work we study the implementation of DTQW on IBM quantum computers and simulators, for the most common topologies found in the literature, i.e. the line, cycle, hypercube, and complete graphs. We will analyze how to theoretically build the circuits based on the shunt decomposition method, how to optimize these circuits using different techniques, how to implement them on IBM quantum computers and simulators, and, finally, as the decompositions are not unique, we will compare their performance in order to choose the decomposition that behaves best and is therefore suitable to be scaled for a larger number of qubits.

The structure of this paper is the following: In section \ref{section2} we introduce the theoretical basis of DTQW and the shunt decomposition method, the method we use to build evolution operators, as block diagonal unitary operators. We then show, in section \ref{section3}, the general way in which block diagonal unitary operators can be mapped into quantum circuit representation in order to be implemented in quantum computers. In section \ref{section4} we apply the method described in section \ref{section3} for the $n$-line, the $n$-cycle, the $n$-hypercube, and the $2^n$-complete graph. We also provide a synthesis for some circuits based on a matrix analysis approach of their evolution operators, and on different techniques to reduce networks of multi-control gates. In section \ref{section5} we implement the quantum circuits derived in section \ref{section4} on IBM quantum computers and Qiskit simulators. In this section we also provide the experimental distributions of the quantum states after measurement, and compare them with the theoretical ones. Finally, in section \ref{conclusion_section}, we present the conclusions 

\section{\label{section2}Theory of Discrete-Time Quantum Walks}

A Discrete-Time Quantum Walk, considers a walker that moves between the vertices of a graph $\mathcal{G}(V,A)$ with adjacency matrix $\mathcal{A}$, where $V$ and $A$ are the vertex and arc sets, respectively. A DTQW is defined by three elements: the quantum state of a walker, the evolution operator of the system and a set of measurement operators. The state of the walker at time $t$ is given by the composite quantum state presented in Eq. \eqref{quantum_state_walker}:

\begin{equation}
\label{quantum_state_walker}
|\psi(t) \rangle = \sum\limits_{i,j} a_{ij}|c_i \rangle \otimes |v_j\rangle
\end{equation}
where $|v_j\rangle \in \mathcal{H}_P$ is a state associated to vertex $v_j \in \mathbb{Z}$, and $|c_k\rangle \in \mathcal{H}_C$, with $c_k \in \mathbb{Z}$, is a state associated to a set of arcs $(v_j,v_i) \in A$. The relation of  relation coin states and arcs of $\mathcal{G}$ is explained in the next paragraphs (see Eq. \eqref{shift_per_diag}). $\mathcal{H}_P$ and $\mathcal{H}_C$ are complex Hilbert spaces of size $n'$ and $m'$, and are called position and coin spaces, respectively. Consider $|v_j\rangle$ and $|c_k\rangle$ to be represented by the corresponding canonical basis. The evolution operator $U$ is given by the product of the shift and coin operators, i.e.

\begin{equation}
    U=S(C\otimes I_{n'})
\end{equation}

One step of the walker consists in the application of $U$ to $|\psi(t)\rangle$, the walker standing on vertex $v_j \in V$ first gets in a superposition of coin states by the action of $C\otimes I_{n'}$, and then moves toward all the adjacent vertices at the same time, by the action of $S$. The state of the system after $t$ steps is given by $|\psi(t)\rangle = U^t |\psi_0 \rangle$, where $|\psi_0 \rangle$ is the initial state of the walker. 

Finally, we define the measurement operators of the system as 
\begin{equation}
\label{measurement_operator}
    M_j = I_{m'} \otimes |v_j\rangle \langle v_j| 
\end{equation}
and the probability to find a walker on vertex $v_j$ after $t$ steps is given by
\begin{equation}
\label{probability_vertex}
    P(|v_j\rangle) = \langle \psi(t)|M_j^{\dagger}M_j|\psi(t) \rangle
\end{equation}

Different methods to construct evolution operators to perform walks on different topologies have been proposed. Particularly, the case in which the shift operator of a DTQW is a block diagonal matrix, whose block elements are permutation matrices that allows us to easily manipulate $S$. Godsil and Zhan \cite{GODSIL2019181} named this method $shunt\; decomposition$, and studied its application to perform DTQWs on regular graphs. Montanaro \cite{montanaro_2007} went further and proved the method to work for strongly connected digraphs, i.e. digraphs in which every pair of vertices are connected by a path, although it also works for graphs with strongly connected subgraphs that do not necessarily connect between them, as will be presented in a particular case later in this paper.

Now we explain the construction of the evolution operator. Let $\mathcal{G}$ be $m'$-regular graph, with adjacency matrix $\mathcal{A}(\mathcal{G})$. Suppose $\mathcal{A}^{\intercal}$ can be decomposed as the sum of $m'$ permutation matrices, $\mathcal{P}^{\intercal}_i \in \mathbb{R}^{n' \times n'}$, i.e.

\begin{equation}
    \label{additive_decomposition_eq}
    \mathcal{A}^{\intercal}=\mathcal{P}^{\intercal}_0+\mathcal{P}^{\intercal}_1+\dots+\mathcal{P}^{\intercal}_{m'-1}
\end{equation}
We call each permutation matrix a $shunt$. Next, we associate each shunt with a coin basis state, and define the shift operator of the system as expressed in Eq. \eqref{shift_per_diag}. 

\begin{equation}
    \label{shift_per_diag}
    S= \sum\limits_{i=0}^{m'-1} |c_i\rangle \langle c_i| \otimes \mathcal{P}^{\intercal}_i
\end{equation}
Thus, a walker with coin state $|c_i\rangle$ will only be able to move through arcs associated with shunt $\mathcal{P}^{\intercal}_i$. Notice that we use matrices $\mathcal{P}^{\intercal}_i$ to construct $S$, since otherwise the quantum walker would move in the opposite direction to the arcs of $\mathcal{G}$.

In explicit matrix notation, the shift operator takes the following form

\begin{equation}
\label{block_diag_shift}
S=
\begin{pmatrix}
\mathcal{P}^{\intercal}_0 & 0 & \dots & 0 \\
0 & \mathcal{P}^{\intercal}_1 & \dots & 0 \\
\vdots & \vdots & \ddots & \vdots \\
0 & 0 & \dots & \mathcal{P}^{\intercal}_{m'-1} 
\end{pmatrix}
\end{equation}
Because $S$ is a block diagonal matrix with unitary matrices as entries, $S$ is unitary too, and contains the same information of the connections between nodes than the original adjacency matrix.

The evolution operator can be completed by the choice of any coin operator $C$ of dimension $m' \times m'$. Nevertheless, in this work we will focus on the study of the circuit implementation of shift operators, and thus we decided to always use Grover and Hadamard operators for this task.

Finally, the evolution operator is applied on a bipartite state $|\psi (t) \rangle \in H_C\otimes H_P$, with $dim(H_C)=m'$ and $dim(H_P)=n'$, and the measurement operators act as defined in Eqs. \eqref{measurement_operator} and (\ref{probability_vertex}).

\section{\label{section3} Mapping of DTQWs to the Quantum Circuit Model}

Let $\mathcal{G}(V,A)$ be a graph with connected components of order $n'=2^n$, whose adjacency matrix can be decomposed into the sum of $m'=2^m$ permutation matrices and whose vertices are labeled by integers. Associated to the vertices of the graph $\mathcal{G}$, there exists a set of $2^n$ canonical basis vectors $\{|v_j \rangle: v_j \in \mathbb{Z}\}$ that span $H_P$, and associated to the arcs connected to vertex $v_i$ there is a set of $2^m$ canonical basis vectors $\{|c_i \rangle: c_i \in \mathbb{Z}\}$ that span $H_C$.

In order to map the basis states of $H_c$ and $H_p$ into a composite state of qubits, we must map $v_j$ and $c_k$ into bitstring notation. Thus, we define the function $f_V: \mathbb{Z} \rightarrow B$, where $B$ is the set of all bitstrings, and consider the basis states of a qubit to be given by

\begin{equation}
|0\rangle =
  \begin{pmatrix}
1 \\
0
  \end{pmatrix}
, \;
|1\rangle =
  \begin{pmatrix}
0 \\
1 
  \end{pmatrix}   
\end{equation}
In this way, we are able to do the following map
\begin{equation}
|f_V(v_i) \rangle = |q_n\dots q_1q_0\rangle = |q_n \rangle  \otimes \dots \otimes |q_1 \rangle \otimes |q_0\rangle
\end{equation}

\noindent
where $q_i \in \{0,1\}$.

This means that every basis vector $|v_i\rangle$ can be rewritten as a composite state given by the tensor product of $n$ qubits. The same mapping is done for the coin space, allowing us to express the state vector of a quantum walker, $|\psi \rangle = |c_i\rangle \otimes |v_i \rangle$, in bitstring notation. 

This mapping can be represented in an $(n+m)$-qubit quantum circuit, where the first $n$ qubits form a register $|q_{n-1} \dots q_1 q_0 \rangle$ that corresponds to the position states, and the last $m$ qubits form a register $|q_{n+m-1} \dots q_{n+1} q_{n} \rangle$ that corresponds to the coin states. Note that the leftmost state is associated with the less significant bit in bitstring representation.

%Little introduction to quantum circuits: Explain how to construct a quantum circuit and what are the parallel lines in a quantum circuit

To completely map a DTQW into a quantum circuit, we need to express the evolution operator, $U=S(C \otimes I_P)$, as a set of quantum gates. A general method can be followed for graphs whose adjacency matrices can be decomposed in $2^m$ permutation matrices $\mathcal{P}^{\intercal}_i$ of size $2^n$, as in Eq. \eqref{additive_decomposition_eq}, which will be used as block diagonal elements of $S$.
\begin{figure}[]
\centering
\subfigure[]{
\includegraphics[scale=0.85]{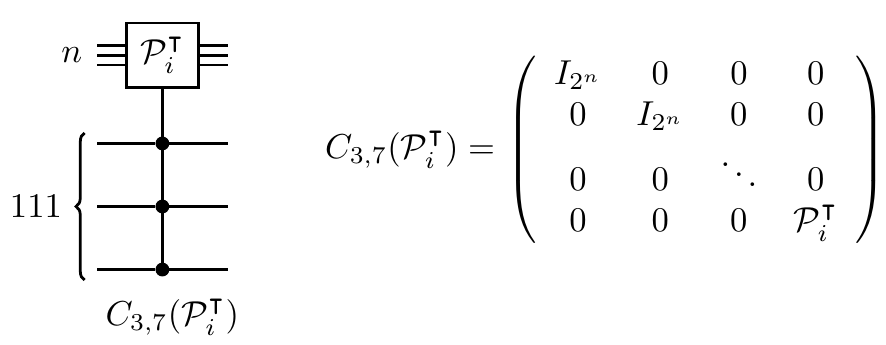}
\label{example_multi-control_U_a}
}

\subfigure[]{
\includegraphics[scale=0.85]{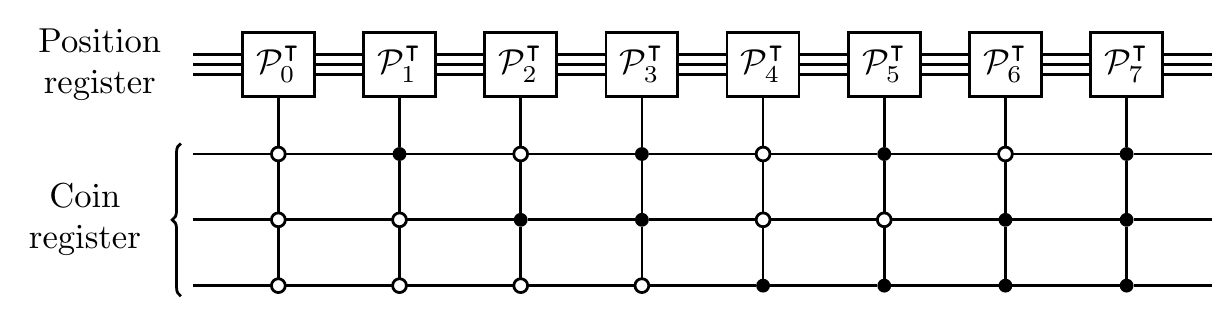}
\label{example_multi-control_U_b}
}
\caption{(a) Example of a multi-control-$\mathcal{P}^{\intercal}_i$ gate, along with its matrix representation. The pattern of white and black dots can be interpreted as binary code, and defines the position of $\mathcal{P}^{\intercal}_i$ in the matrix representation of $C_{3,7}(\mathcal{P}^{\intercal}_i)$, which follows the form of Eq. \eqref{block_diag_shift}. (b) Circuit implementation of a block diagonal shift operator. Notice the controls and the target of each gate span the coin and position registers, respectively, which holds whenever mapping a block diagonal operator to circuit form}
\label{example_multi-control_U}
\end{figure}
Given that all matrices $\mathcal{P}^{\intercal}_i$ are unitary, there exists a quantum gate representation for all of them. Out of the quantum gate $\mathcal{P}^{\intercal}_i$, we can create a controlled gate, $C_{m, j}(\mathcal{P}^{\intercal}_i)$, which uses all the qubits of the coin register as control qubits, as shown in Fig. \ref{example_multi-control_U_a}. We call the black and the white dots in this figure $controls$. The pattern that they create can be interpreted as binary code, where the black controls represent a 1, the white controls represent a 0 and the less significant bit of the bitstring is the uppermost control in the position register. The second subindex $j = 0,1, \dots, 2^m-1$ in $C_{m, j}(\mathcal{P}^{\intercal}_i)$ represents the value of the bitstring formed by the control qubits of the gate, while $m$ represents the number of control qubits. Notice that we reserve this notation for fully control gates which use the qubits at the bottom of the target gate as control qubits. If the target gate is controlled using the top qubits we use the notation $C^{r,k}(U)$. In the case the gate is fully controlled using both top and bottom qubits, we use the notation $C^{r,k}_{s,l}(U)$. 

The result of controlling $\mathcal{P}^{\intercal}_i$ is a $2^{n+m}$ block diagonal matrix, where all the diagonal elements of the matrix are the $2^n \times 2^n$ identity, $I_{2^n}$, except for the $j$th element, which will be the matrix $\mathcal{P}^{\intercal}_i$. This is given by Eq. \eqref{controlled-M_i}:
\begin{equation}
    \label{controlled-M_i}
    C_{m, j}(\mathcal{P}^{\intercal}_i) = I_{2^{jn}} \oplus \mathcal{P}^{\intercal}_{i}\oplus I_{2^{(2^m-j-1)n}}
\end{equation}
where $i,j = 0,1, \dots, 2^{m}-1$ and we define $I_0$ as a zero-dimensional matrix.. Notice that $i$ and $j$ do not have to coincide, which just means that the matrices $\mathcal{P}^{\intercal}_i$ can be shuffled in the diagonal indistinguishably.

This way, we can find $2^m$ block diagonal gates $C_{m,j}(\mathcal{P}^{\intercal}_{i})$ in which the matrices $\mathcal{P}^{\intercal}_{i}$ are placed in a different position $j$ relative to each other, and the rest of the elements are $I_{2^n}$. When these controlled gates are placed next to each other in a quantum circuit, as in Fig. \ref{example_multi-control_U_b}, the associated matrix representation is given by

\begin{equation}
    \label{general_shift_control_product}
    S = \prod\limits_{j=0}^{2^m-1} C_{m, j}(\mathcal{P}^{\intercal}_{i})
\end{equation}
a block diagonal operator whose elements are the additive decomposition of the adjacency matrix associated to graph $\mathcal{G}$, i.e. we obtain Eq. \eqref{block_diag_shift}. 

To complete the mapping of the evolution operator we can add a Hadamard coin by simply placing a single-qubit Hadamard gate to all the qubits of the coin register to the left of the circuit for the shift operator.

\section{\label{section4}Implementation of Shift Operators for DTQWs on Common Topologies}

In this section, we will analyze the shunt decomposition of the adjacency matrices of the most common graphs studied in the field of quantum walks: the line with n vertices, the $n$-cycle graph, the $2^n$-hypercube and the $2^n$-complete graph with self-loops, as well as their quantum circuit implementation. 
\subsection{DTQW on the n-cycle graph}
\label{cycle_graph_sec}
The adjacency matrix of a $2^n$-cycle graph can be deduced following the pattern presented in \cite{carnia_adjacency_clycle_complete}, as shown in Eq. \eqref{cicle_adj_matrix}.

\begin{equation}
\mathcal{A}_{cycle} =
\label{cicle_adj_matrix}
\begin{pmatrix}
0 & 1 & 0 & 0 & \dots & 0 & 0 & 0 & 1 \\
1 & 0 & 1 & 0 & \dots & 0 & 0 & 0 & 0 \\
0 & 1 & 0 & 1 & \dots & 0 & 0 & 0 & 0 \\
0 & 0 & 1 & 0 & \dots & 0 & 0 & 0 & 0 \\
\vdots & \vdots & \vdots & \vdots & \ddots & \vdots & \vdots & \vdots & \vdots \\
0 & 0 & 0 & 0 & \dots & 0 & 1 & 0 & 0 \\
0 & 0 & 0 & 0 & \dots & 1 & 0 & 1 & 0 \\
0 & 0 & 0 & 0 & \dots & 0 & 1 & 0 & 1 \\
1 & 0 & 0 & 0 & \dots & 0 & 0 & 1 & 0 \\
\end{pmatrix}
\end{equation}

\begin{figure}[h!]
\centering
\subfigure[]{
\includegraphics[scale=1]{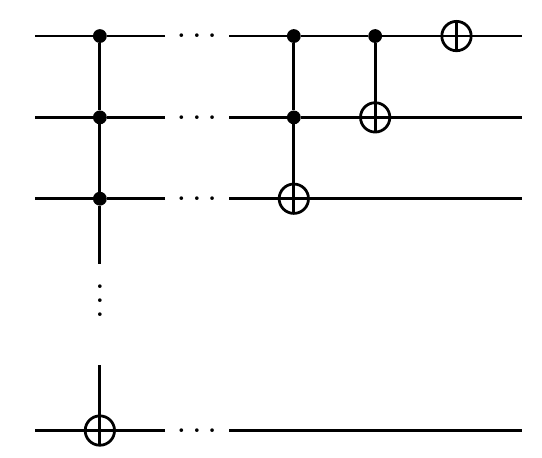}
\label{inc_and_dec_gates_a}
}
\subfigure[]{
\includegraphics[scale=1]{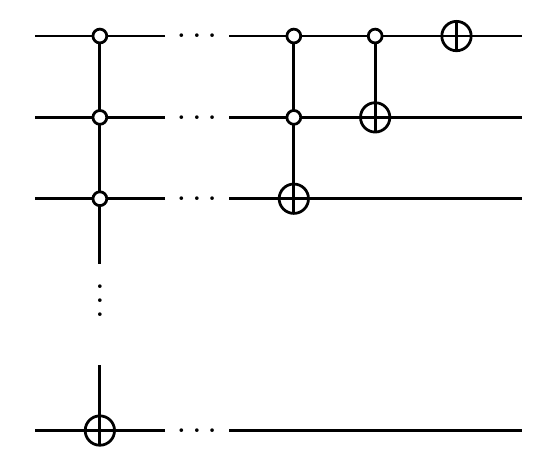}
\label{inc_and_dec_gates_b}
}
\caption{(a) $n$-qubit increment gate. (b) $n-$qubit decrement gate}
\label{inc_and_dec_gates}
\end{figure}

In \cite{li_yang_increment}, Li {\it et \ al.} present general sequences of multi-control not gates that conform $n$-qubit increment and decrement gates, as shown in Fig. (\ref{inc_and_dec_gates}). The transpose of the matrix forms of these gates are given in Eqs. \eqref{incement__matrix} and \eqref{decrement_matrix}.

%\begin{figure}[h!]
%  \includegraphics[scale=0.45]{inc_and_dec_gates.JPG}
%  \caption{a) n-qubit increment gate. b) n-qubit decrement gate.}
%  \label{inc_and_dec_gates}
%\end{figure}

\begin{equation}
\mathcal{A}^{\intercal}_{inc}=
\label{incement__matrix}
\begin{pmatrix}
0 & 0 & 0 & 0 & \dots & 0 & 0 & 0 & 1 \\
1 & 0 & 0 & 0 & \dots & 0 & 0 & 0 & 0 \\
0 & 1 & 0 & 0 & \dots & 0 & 0 & 0 & 0 \\
0 & 0 & 1 & 0 & \dots & 0 & 0 & 0 & 0 \\
\vdots & \vdots & \vdots & \vdots & \ddots & \vdots & \vdots & \vdots & \vdots \\
0 & 0 & 0 & 0 & \dots & 0 & 0 & 0 & 0 \\
0 & 0 & 0 & 0 & \dots & 1 & 0 & 0 & 0 \\
0 & 0 & 0 & 0 & \dots & 0 & 1 & 0 & 0 \\
0 & 0 & 0 & 0 & \dots & 0 & 0 & 1 & 0 
\end{pmatrix}
\end{equation}

\begin{equation}
\mathcal{A}^{\intercal}_{dec} =
\begin{pmatrix}
0 & 1 & 0 & 0 & \dots & 0 & 0 & 0 & 0 \\
0 & 0 & 1 & 0 & \dots & 0 & 0 & 0 & 0 \\
0 & 0 & 0 & 1 & \dots & 0 & 0 & 0 & 0 \\
0 & 0 & 0 & 0 & \dots & 0 & 0 & 0 & 0 \\
\vdots & \vdots & \vdots & \vdots & \ddots & \vdots & \vdots & \vdots & \vdots \\
0 & 0 & 0 & 0 & \dots & 0 & 1 & 0 & 0 \\
0 & 0 & 0 & 0 & \dots & 0 & 0 & 1 & 0 \\
0 & 0 & 0 & 0 & \dots & 0 & 0 & 0 & 1 \\
1 & 0 & 0 & 0 & \dots & 0 & 0 & 0 & 0 \\
\end{pmatrix}
\label{decrement_matrix}
\end{equation}
Thus, the transpose adjacency matrix of a $2^n$-cycle graph can be written as a linear combination of the $n$-qubit transpose increment and decrement operators

\begin{equation}
    \mathcal{A}^{\intercal}_{cycle} = \mathcal{A}^{\intercal}_{inc}+\mathcal{A}^{\intercal}_{dec}
\end{equation}
where both $\mathcal{A}^{\intercal}_{inc}$ and $\mathcal{A}^{\intercal}_{dec}$ are unitary. 

\begin{figure}[b!]
\centering
\includegraphics[scale=1]{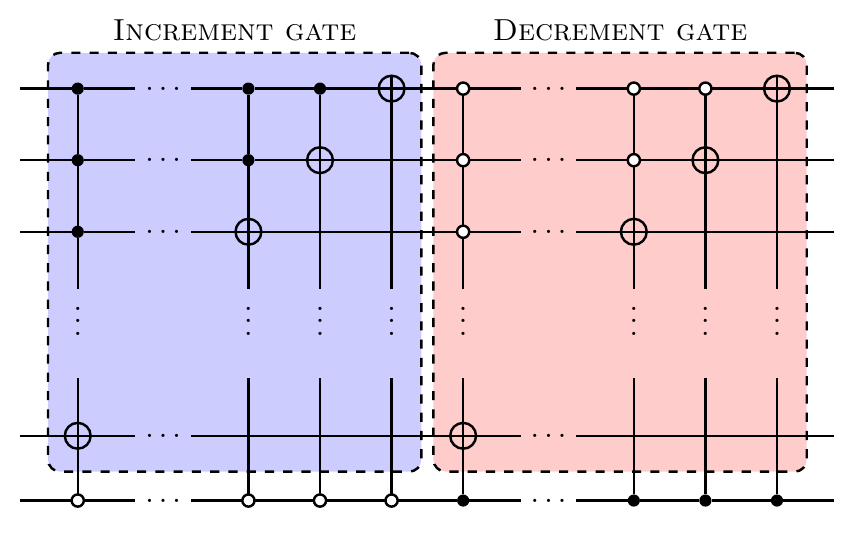}
\caption{Shift operator for a $2^n$-cycle graph composed of an $n-$qubit-controlled increment gate and an $n-$qubit-controlled decrement gate}
\label{usual_shift_line}
\end{figure}

According to Eq. \eqref{block_diag_shift}, once we have the additive decomposition of an adjacency matrix associated to a graph $\mathcal{G}$ into unitary matrices, we can construct a shift operator for a DTQW as shown in Eq. \eqref{shift_2n-cycle}.

\begin{equation}
S =
\label{shift_2n-cycle}
\begin{pmatrix}
\mathcal{A}^{\intercal}_{inc} & 0 \\
0 & \mathcal{A}^{\intercal}_{dec}
\end{pmatrix}
\end{equation}
Moreover, according to Eq. \eqref{general_shift_control_product} we can build a sequence of controlled gates such that when applied subsequently, they give out Eq. \eqref{shift_2n-cycle}. The sequence of gates $C_{1,0}(\mathcal{A}^{\intercal}_{inc})$ and $C_{1,1}(\mathcal{A}^{\intercal}_{dec})$ is presented in Fig. (\ref{usual_shift_line}). This is the common implementation of a circuit for a cycle graph, although if a further matrix approach is made to Eq. \eqref{shift_2n-cycle} the number of gates can be reduced in half.
Firstly, notice that 

\begin{equation}
    \mathcal{A}_{dec} = \mathcal{A}^{\intercal}_{inc}
\end{equation}
Now we define the following $2^n \times 2^n$ matrix

\begin{equation}
J =
\label{second_diagonal_matrix}
\begin{pmatrix}
0 & 0 & 0 & 0 & \dots & 0 & 0 & 0 & 1 \\
0 & 0 & 0 & 0 & \dots & 0 & 0 & 1 & 0 \\
0 & 0 & 0 & 0 & \dots & 0 & 1 & 0 & 0 \\
0 & 0 & 0 & 0 & \dots & 1 & 0 & 0 & 0 \\
\vdots & \vdots & \vdots & \vdots & \ddots & \vdots & \vdots & \vdots & \vdots \\
0 & 0 & 0 & 1 & \dots & 0 & 0 & 0 & 0 \\
0 & 0 & 1 & 0 & \dots & 0 & 0 & 0 & 0 \\
0 & 1 & 0 & 0 & \dots & 0 & 0 & 0 & 0 \\
1 & 0 & 0 & 0 & \dots & 0 & 0 & 0 & 0 \\
\end{pmatrix}
\end{equation}
When $J$ is applied simultaneously to the right and to the left of a circulant matrix, i.e. a matrix of the form presented in Eq. \eqref{circulant_matrix}, the result is the transpose of the original matrix \cite{shakeel_2020, golub_2013}. 

\begin{equation}
C=
\begin{pmatrix}
c_0 & c_{n-1} & \dots & c_2 & c_1 \\
c_1 & c_0 & c_{n-1} & \dots & c_2 \\
\vdots & c_1 & c_0 & \ddots & \vdots \\
c_{n-2} & \vdots & \ddots & \ddots & c_{n-1} \\
c_{n-1} & c_{n-2} & \dots & c_1 & c_0
\end{pmatrix}
\label{circulant_matrix}
\end{equation}
Given that $\mathcal{A}^{\intercal}_{inc}$ lies on this category, $\mathcal{A}^{\intercal}_{dec}$ can be written alternatively as

\begin{equation}
    \mathcal{A}^{\intercal}_{dec} = J \mathcal{A}^{\intercal}_{inc} J
\end{equation}
so that (\ref{shift_2n-cycle}) becomes

\begin{equation}
S =
\label{shift_2n-cycle_J_mat}
\begin{pmatrix}
\mathcal{A}^{\intercal}_{inc} & 0 \\
0 & J \mathcal{A}^{\intercal}_{inc} J
\end{pmatrix}
\end{equation}
This matrix can be decomposed in the following way

\begin{equation}
S =
\label{shift_2n-cycle_J_mat_dec}
\begin{pmatrix}
I_{2^n} & 0 \\
0 & J
\end{pmatrix}
\begin{pmatrix}
\mathcal{A}^{\intercal}_{inc} & 0 \\
0 & \mathcal{A}^{\intercal}_{inc}
\end{pmatrix}
\begin{pmatrix}
I_{2^n} & 0 \\
0 & J
\end{pmatrix}
\end{equation}
where

\begin{equation}
    I_{2} \otimes \mathcal{A}^{\intercal}_{inc}  =
\label{augmented_inc_gate}
\begin{pmatrix}
\mathcal{A}^{\intercal}_{inc} & 0 \\
0 & \mathcal{A}^{\intercal}_{inc}
\end{pmatrix}
\end{equation}
\begin{figure*}[t!]
\centering
\subfigure[]{
\includegraphics[scale=0.63]{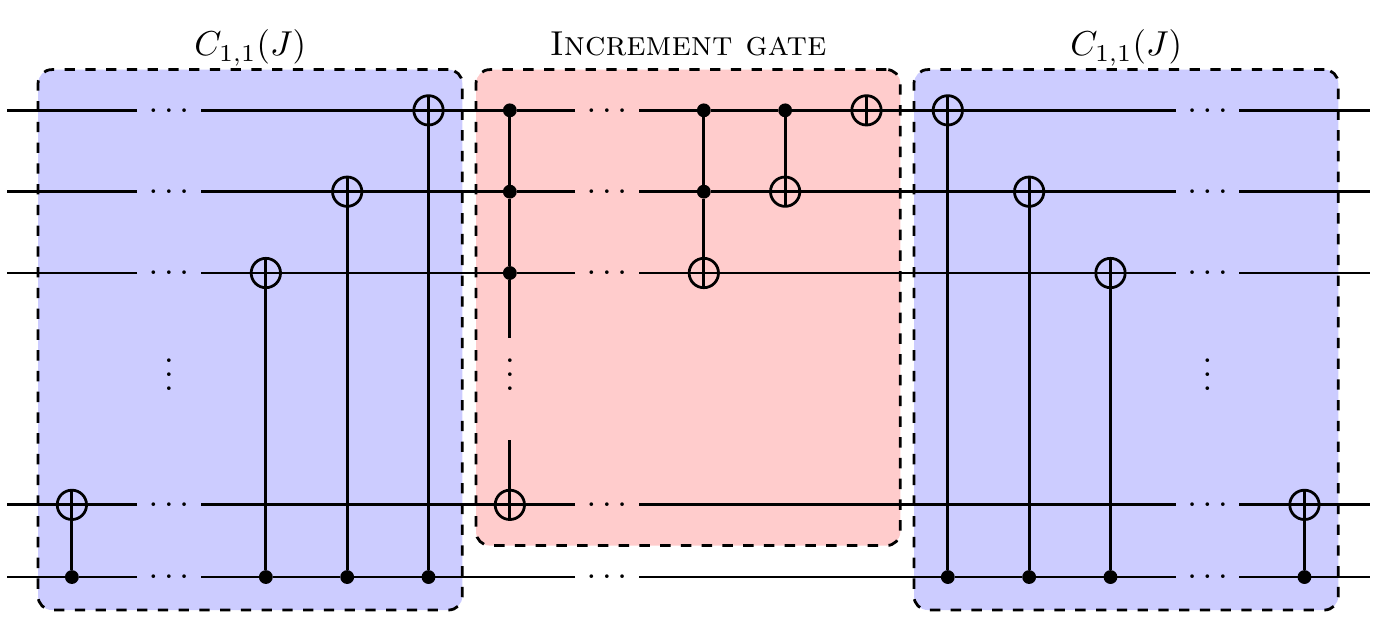}
\label{inc-dec_operator_a}
}
\subfigure[]{
\includegraphics[scale=0.63]{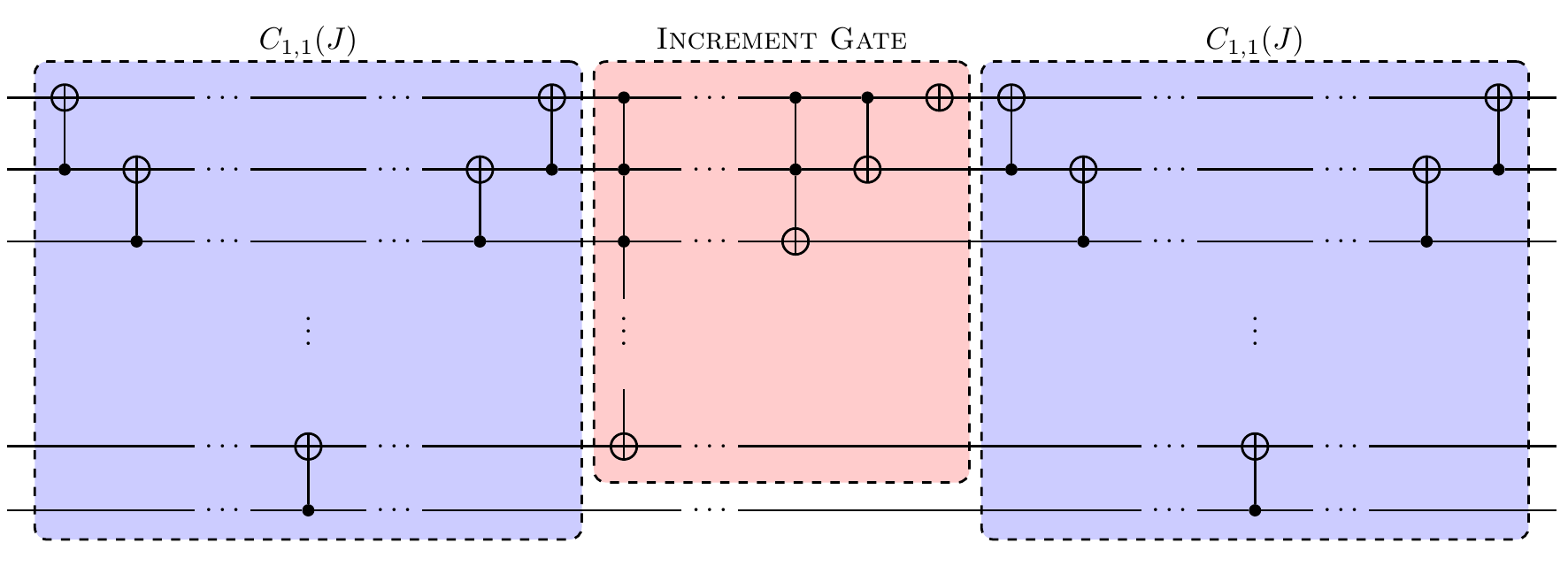}
\label{inc-dec_operator_b}
}
\caption{Both (a) and (b) display a quantum circuit for the $S$ operator of a DTQW on a $2^n$-cycle graph, according to Eq. \eqref{shift_2n-cycle_J_mat_dec}. The gates at the sides of the increment circuit are different implementations of $C_{1,1}(J)$ (Eq. \eqref{controlled-J_1}). The equivalence between both circuit forms of $C_{1,1}(J)$ has been used in \cite{mottonen} and proven in \cite{qc_paulesina}}
\label{inc-dec_operator}
\end{figure*}
and
\begin{equation}
    C_{1,1}(J) =
\label{controlled-J_1}
\begin{pmatrix}
I_{2^n} & 0 \\
0 & J
\end{pmatrix}
\end{equation}
$C_{1,1}(J)$ can be obtained by controlling with a black control a network of $\sigma_x$ gates applied to all qubits of the position register, given that $\sigma_x^{\otimes n}$ generates $J$, or by implementing CNOT gates in cascade configuration, as shown in Fig. (\ref{inc-dec_operator}), which displays two reduced quantum circuit implementations of the shift operator presented in Eq. \eqref{shift_2n-cycle}. The equivalence between the two circuit forms of $C_{1,1}(J)$ was used in \cite{mottonen} to decompose general multi-control gates and proven in \cite{qc_paulesina} for low dimensional instances through a recursion relation which can be used to obtain higher-dimensional instances. 

The circuit presented in Fig. \ref{inc-dec_operator_a} was proposed by \cite{shakeel_2020}. These circuits are suitable to efficiently simulate DTQWs on cycle graphs of size $2^n$ only. 

%In light of this, a further approach based on the matrix analysis of  the shift operator, was done in order to implement quantum circuits capable of simulating a DTQW on a general $n$-cycle graph.

We  now present a further approach, based on the matrix analysis of  the shift operator, in order to implement quantum circuits capable of simulating a DTQW on a general $n$-cycle graph.

First consider the matrix associated to an $n$-quibit increment gate (Eq. \eqref{incement__matrix}). In \cite{unitary_decomposition_li}, Li {\it et al.}  define a general method to decompose a square unitary matrix of order $k$ into a product of no more than $k(k-1)/2$ two-level unitary matrices of the same size. A special case of this decomposition is the one in which the matrices $\mathcal{T}_i$ have size $2^n \times 2^n$ and take the following form

\begin{equation}
\label{switch_two_rows_equation}
\mathcal{T}_{i}=I_{i} \oplus \sigma_x \oplus I_{2^{n}-i-2}; \ i=0,1,...,2^{n}-2
\end{equation}
or in explicit matrix notation

\begin{equation}
\mathcal{T}_i =
\label{two_level_matrix}
\begin{pmatrix}
I_{i} & 0 & 0 \\
0 & \sigma_x & 0 \\
0 & 0 & I_{2^{n}-i-2}
\end{pmatrix},
\end{equation}
where we define $I_0$ as a zero-dimensional matrix.

In simple terms, $\mathcal{T}_i$ is an identity matrix in which a $\sigma_x$ matrix was replaced in the main diagonal, and the subindex $i$ in Eq. \eqref{two_level_matrix} represents the number of ones above $\sigma_x$. The action of $\mathcal{T}_i$ on any matrix is to exchange the rows $i$ and $i+1$. In group theory this operation is known as \textit{adjacent transposition} \cite{rotman_1995}, thus we call $\mathcal{T}_i$ the adjacent transposition operator.

The increment and decrement operators (Eqs. \eqref{incement__matrix} and \eqref{decrement_matrix}) can be decomposed as the product of all matrices $\mathcal{T}_i$ in descending and ascending order, respectively, i.e.

\begin{equation}
    \mathcal{A}^{\intercal}_{inc} = \mathcal{T}_0 \mathcal{T}_1 \cdots \mathcal{T}_{2^n-2}
\end{equation}

\begin{equation}
    \mathcal{A}^{\intercal}_{dec} = \mathcal{T}_{2^n-2} \cdots  \mathcal{T}_1 \mathcal{T}_0
\end{equation}
which can be though of as applying the sequence of $\mathcal{T}_i$ matrices to the identity $I_{2^n}$ in order to switch its rows.

Now, if instead of multiplying the whole sequence of matrices $\mathcal{T}_i$ to obtain the $n$-qubit transpose increment operator, we truncate the product up to the first $k$ matrices, i.e., we obtain the operator of Eq. \eqref{inc_ak}.

\begin{equation}
\label{inc_ak}
(\mathcal{A}^k_{inc})^{\intercal} = \mathcal{T}_0\mathcal{T}_1\dots\mathcal{T}_{k-2}
\end{equation}
or explicitly

\begin{equation}
(\mathcal{A}^k_{inc})^{\intercal} =  (\mathcal{A}'_{inc})^{\intercal} \oplus I_{2^n-k} =
\begin{pmatrix}
(\mathcal{A}'_{inc})^{\intercal} & 0 \\
0 & I_{2^n-k}
\end{pmatrix}
\label{ak_dec}
\end{equation}
where $(\mathcal{A}'_{inc})^{\intercal}$ is the $k \times k$ transpose increment operator, i.e. $(\mathcal{A}'_{inc})^{\intercal}$ is a $k \times k$ version of $\mathcal{A}^{\intercal}_{inc}$ in Eq. (\ref{incement__matrix}), which is always $2^n \times 2^n$, but with $k \le 2^n$. Thus, although $(\mathcal{A}^k_{inc})^{\intercal}$ is a $2^n \times 2^n$ operator, when applied to a register of $n$ qubits, it will only have effect on the first $k$ states, leaving the rest intact.

\begin{figure}[h!]
\centering
\subfigure[]{
\includegraphics[scale=0.8]{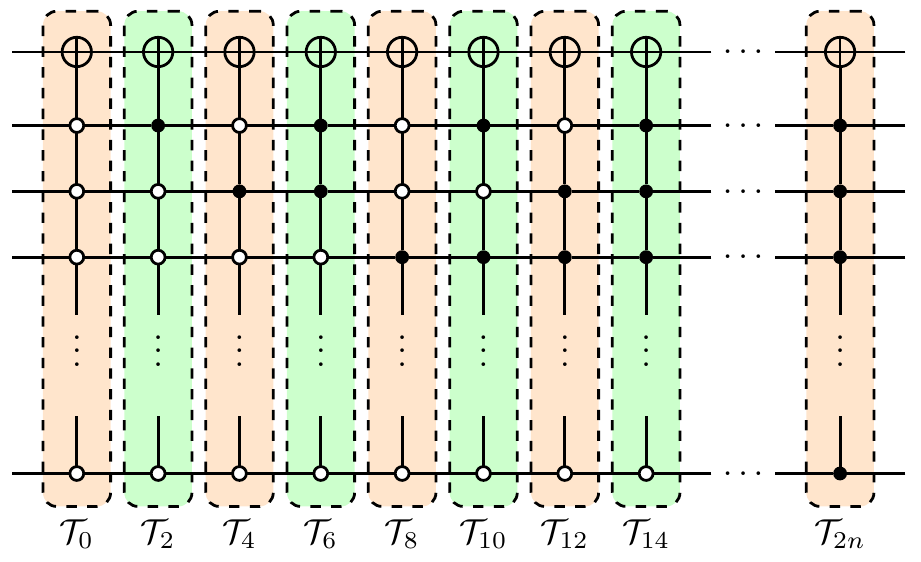}
\label{seq_2j_4j+1_a}
}
\subfigure[]{
\includegraphics[scale=0.8]{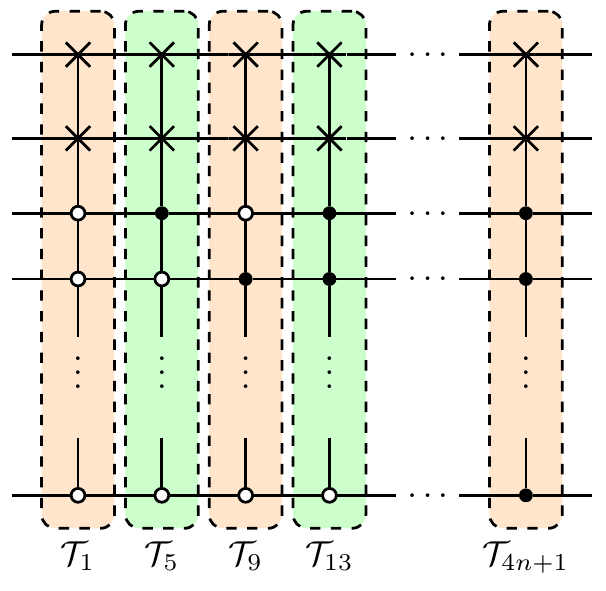}
\label{seq_2j_4j+1_b}
}
\caption{(a) Displays the sequence to create gates $\mathcal{T}_i$ whose index follow the relation $i = 2j$. (b) Displays the sequence to create gates $\mathcal{T}_i$ whose index follow the relation $i = 4j+1$. Both sequences are given by all the combinations of multi-controlled not or swap gates}
\label{seq_2j_4j+1}
\end{figure}

%\newpage{}
The same can be done for the transpose decrement operator, i.e. we define $(\mathcal{A}^k_{dec})^{\intercal} = \mathcal{T}_{k-2} \cdots \mathcal{T}_{1} \mathcal{T}_{0}$. Explicitly 

\begin{equation}
(\mathcal{A}^k_{dec})^{\intercal} = (\mathcal{A}'_{dec})^{\intercal} \oplus I_{2^n-k} =
\begin{pmatrix}
(\mathcal{A}'_{dec})^{\intercal} & 0 \\
0 & I_{2^n-k}
\end{pmatrix}
\label{ak_inc}
\end{equation}
%where $(\mathcal{A}'_{dec})^{\intercal}$ is the $k \times k$ decrement operator. 

The circuit implementation of all adjacent transposition operators $\mathcal{T}_i$ is not trivial, given that the quantum gates corresponding to operators with consecutive indices do not follow the same structure. We present three quantum gate structures that generate transposition operators $\mathcal{T}_i$ for which $i$ belongs to one of the sequences $(2j)_{j=0}^{j=2^{n-1}}$, $(4j+1)_{j=0}^{j=2^{n-2}}$, and $(4j+3)_{j=0}^{j=2^{n-2}-1}$. The union of all the sequences generate all natural numbers from $0$ to $2^n-2$, thus the combination of transposition operators corresponding to different sequences generate every $\mathcal{T}_i$ that follows Eq. \eqref{switch_two_rows_equation}. Figures \ref{seq_2j_4j+1_a} and \ref{seq_2j_4j+1_b} show the general pattern of transposition operators associated to the first and second sequences, respectively. Each gate uses all the qubits of the position register for its construction. The transposition operators associated to the third sequence must follow a more complex process for their construction that will be described next.

Firstly, the transposition gates associated to $(4j+3)_{j=0}^{j=2^{n-2}-1}$ will only be used in the case where the position register consists of three or more qubits, otherwise the gates associated to the first two sequences alone will suffice, given that for $n$ qubits we can have a maximum of $2^n-1$ transposition gates. Now, to construct transposition gates for position registers of more than two qubits, we will make use of multi-control SWAP gate whose controls are placed in the upper qubits, and we will denote it by $C^{r,i}(SWAP)$, where we remark that $r$ and $i$ refer to the number of controls and the value of the binary string formed by the controls, respectively, as described in section \ref{section3}. 

\begin{figure*}[t!]
\centering
\subfigure[]{
\includegraphics[scale=0.55]{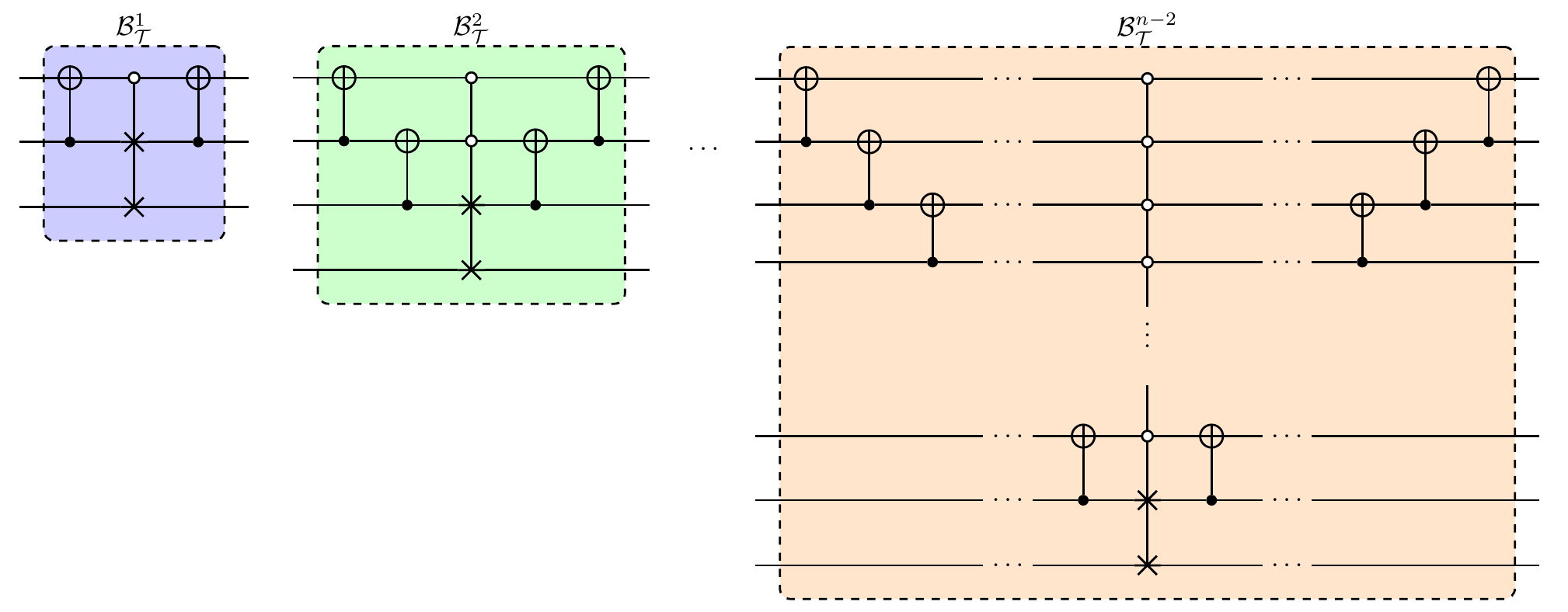}
\label{sequence_4j+3_a}
}
\subfigure[]{
\includegraphics[scale=0.45]{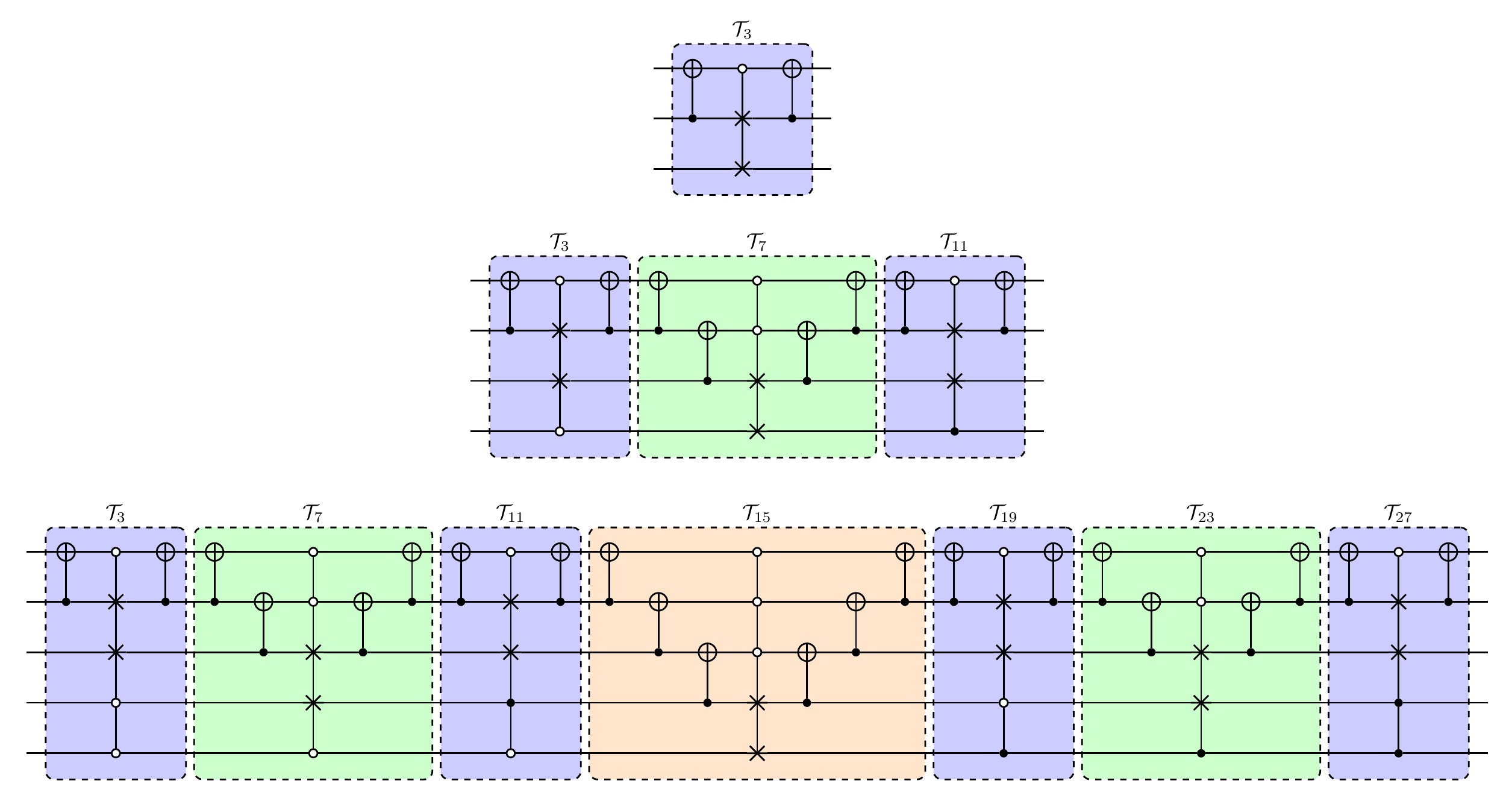}
\label{sequence_4j+3_b}
}
\caption{(a) Sequence of basis gates $\mathcal{B}_\mathcal{T}^r$ to generate the set of transposition gates $\mathcal{T}_i$ associated to $(4j+3)_{j=0}^{j=2^{n-2}-1}$ as the number of qubits increaces. (b) Sequence of circuits that contain all the transposition gates $\mathcal{T}_i$ associated to $(4j+3)_{j=0}^{j=2^{n-2}-1}$ for $n = 3, 4, 5$. This sequence allows us to see that as we increase the number of qubits, the transposition gates for $n$ qubits can be constructed using the set of gates for $n-1$ qubits, i.e. if $\mathcal{S}_\mathcal{T}^{n}$ is the set of transposition gates associated to $(4j+3)_{j=0}^{j=2^{n-2}-1}$ for $n$ qubits, then the recurrence relation $\mathcal{S}_\mathcal{T}^{n}=C_{1,0}(\mathcal{S}_\mathcal{T}^{n-1})\mathcal{B}_\mathcal{T}^{n-2}C_{1,1}(\mathcal{S}_\mathcal{T}^{n-1})$ complies
}
\label{sequence_4j+3}
\end{figure*}

The $C^{r,i}(SWAP)$ gate will allow us to define a set of basis gates which will serve to generate different transposition gates $\mathcal{T}_i$ through a recurrence relation. Each basis gate, $\mathcal{B}_\mathcal{T}^r$, is composed of a $C^{r,0}(SWAP)$ gate at the core, surrounded to the left and to the right by CNOT gates in descending and ascending echelon-like form, respectively, as can be seen in the sequence presented in Fig. \ref{sequence_4j+3_a}. The superindex in $\mathcal{B}_\mathcal{T}^r$ stands for the number of controls in the core $C^{r,0}(SWAP)$ gate. For $n \ge 3$ qubits, there will be $n-2$ basis gates following the pattern of Fig. \ref{sequence_4j+3_a}

Once we have constructed all basis gates $\mathcal{B}_\mathcal{T}^r$ for $n \ge 3$ qubits, the set of transposition gates $\mathcal{T}_i$ associated to $(4j+3)_{j=0}^{j=2^{n-2}-1}$ can be obtained by a recursive relation. To explain how this recursive relation works, we refer to Fig. \ref{sequence_4j+3_b} to explain the first instances of the relation and then provide a general rule.

The upper circuit in Fig. \ref{sequence_4j+3_b} takes place in a $3$-qubit register, which means there is place for one transposition gate, thus the transposition gate $\mathcal{T}_3$ coincides with the basis gate $\mathcal{B}_\mathcal{T}^1$. The middle circuit in this figure takes place in a $4$-qubit register, thus there is place for $3$ transposition gates. The center gate, $\mathcal{T}_7$, coincides with the basis gate $\mathcal{B}_\mathcal{T}^2$, and the gates to left and right of $\mathcal{T}_7$ are built by taking the $3$-qubit $\mathcal{T}_3$ gate and controlling the core $C^{1,0}(SWAP)$ gate with a bottom white and a black control, respectively. For the lower circuit, we follow a similar rationale. We place the basis gate $\mathcal{B}_\mathcal{T}^3$ at the center of the sequence, then take the sequence of $4$-qubit $\mathcal{T}_i$ gates built in the previous circuit and place them to the left and right of $\mathcal{B}_\mathcal{T}^3$ but now with an extra bottom white and black control, respectively, to the core $C^{r,i}(SWAP)$ gates associated to $3$-qubit $\mathcal{T}_i$ gates. In general, the sequence of transposition gates associated to $(4j+3)_{j=0}^{j=2^{n-2}-1}$ for a register of $n$ qubits, is obtained by taking the $\mathcal{B}_\mathcal{T}^{n-2}$ basis gate and use it as the center gate of the sequence, then taking the sequence of $\mathcal{T}_i$ gates for a register of $n-1$ qubits, which we will call $\mathcal{S}_\mathcal{T}^{n-1}$ for simplicity, and finally adding a controlled version of $\mathcal{S}_\mathcal{T}^{n-1}$ to the left and right of $\mathcal{B}_\mathcal{T}^{n-2}$, using bottom white and black controls, respectively. In short notation, $\mathcal{S}_\mathcal{T}^{n}=C_{1,0}(\mathcal{S}_\mathcal{T}^{n-1})\mathcal{B}_\mathcal{T}^{n-2}C_{1,1}(\mathcal{S}_\mathcal{T}^{n-1})$. The index of each $n$-qubit $\mathcal{T}_i$ gate is assigned following $(4j+3)_{j=0}^{j=2^{n-2}-1}$ from left to right.

In order to apply the generalization of the increment and decrement operators to $k$ qubits to a DTQW, we will proceed  similarly as in the case of a $2^n$-cycle graph. That is, we build controlled gates out of $(\mathcal{A}^k_{inc})^{\intercal}$ and $(\mathcal{A}^k_{dec})^{\intercal}$, so that we compute the operators presented in Eqs. \eqref{c_cero_uno} and \eqref{c_uno_uno}:

\begin{equation}
    C_{1,0}((A^k_{inc})^{\intercal}) =
\begin{pmatrix}
(\mathcal{A}'_{inc})^{\intercal} & 0 & 0 & 0 \\
0 & I_{2^n-k} & 0 & 0 \\
0 & 0 & I_k & 0  \\
0 & 0 & 0 & I_{2^n-k}
\end{pmatrix}
\label{c_cero_uno}
\end{equation}

%and 

\begin{equation}
    C_{1,1}((\mathcal{A}^k_{dec})^{\intercal}) =
\begin{pmatrix}
I_k & 0 & 0 & 0 \\
0 & I_{2^n-k} & 0 & 0 \\
0 & 0 & (\mathcal{A}^k_{dec})^{\intercal} & 0  \\
0 & 0 & 0 & I_{2^n-k},
\end{pmatrix}
\label{c_uno_uno}
\end{equation}
which act on quantum states $|\psi\rangle$ with coin state $|0\rangle$ and $|1\rangle$, respectively. Thus the shift operator of a DTQW on a $k$-cycle graph is given by $S=C_{1,0}((\mathcal{A}^k_{inc})^{\intercal})C_{1,1}((\mathcal{A}^k_{dec})^{\intercal})$, i.e.

\begin{equation}
\small
S =
\begin{pmatrix}
(\mathcal{A}'_{inc})^{\intercal} & 0 & 0 & 0 \\
0 & I_{2^n-k} & 0 & 0 \\
0 & 0 & (\mathcal{A}'_{dec})^{\intercal} & 0  \\
0 & 0 & 0 & I_{2^n-k}
\end{pmatrix}
\label{k_shift_op}
\end{equation}

\begin{figure}[h!]
\centering
\subfigure[]{
\includegraphics[scale=0.9]{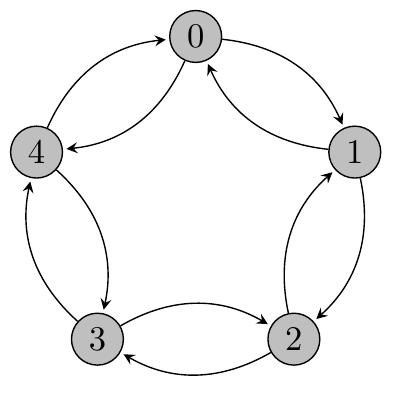}
\label{5-cycle_graph_a}
}
\subfigure[]{
\includegraphics[scale=0.9]{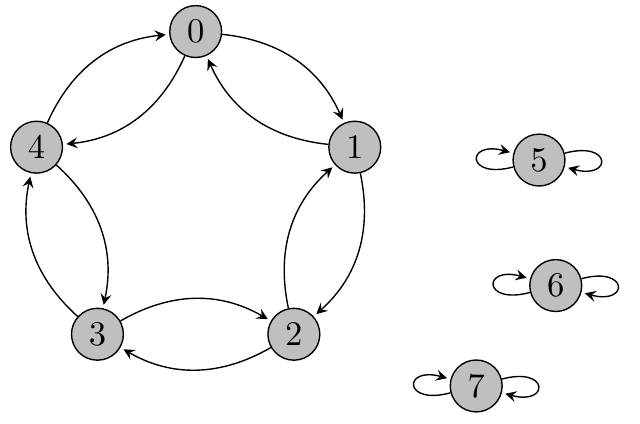}
\label{5-cycle_graph_b}
}
\subfigure[]{
\includegraphics[scale=1]{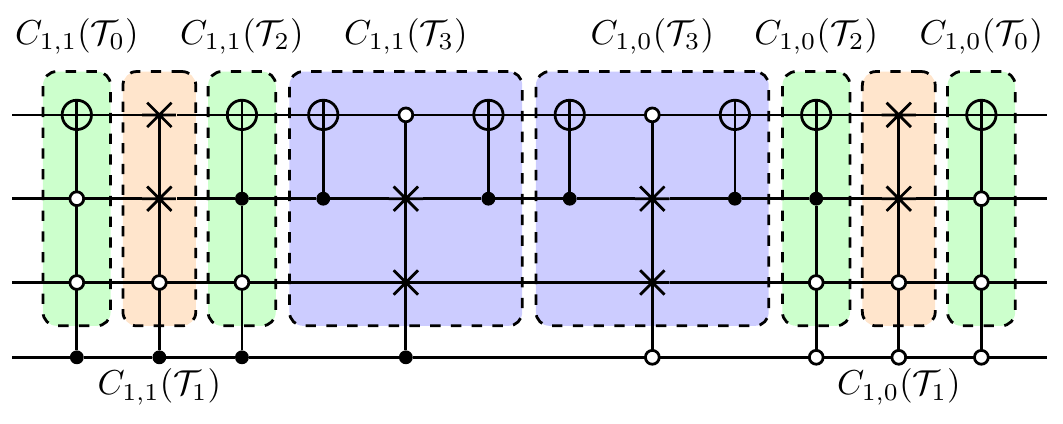}
\label{5-cycle_graph_c}
}
\caption{(a) $5$-cycle graph. (b) 8-vertex graph which contains a 5-cycle graph as a subgraph and three isolated nodes. (c) Circuit implementation of the evolution operator of a DTQW on (b), according to Eq. \eqref{k_shift_op}. To perform a DTQW on the 5-cycle using circuit (c) initialize the position state of the walker on any state $|v\rangle$ for which $v=0,\dots,4$}
\label{5-cycle_graph}
\end{figure}

As an example, consider the 5-cycle graph (see. Fig. \ref{5-cycle_graph_a}), with adjacency matrix which can be decomposed as a sum of $5 \times 5$ increment and decrement operators. 
\begin{equation}
(\mathcal{A}_{5c})^{\intercal} =
\begin{pmatrix}
0 & 1 & 0 & 0 & 1 \\
1 & 0 & 1 & 0 & 0 \\
0 & 1 & 0 & 1 & 0  \\
0 & 0 & 1 & 0 & 1  \\
1 & 0 & 0 & 1 & 0
\end{pmatrix}
\end{equation}
In order to be implementable as a quantum circuit, an operator must have size $2^n \times 2^n$, thus we augment both $5 \times 5$ increment and decrement operators according to equations \eqref{ak_dec} and \eqref{ak_inc}, and we obtain 
\begin{equation}
    (\mathcal{A}^5_{inc})^{\intercal} = 
\begin{pmatrix}
0 & 0 & 0 & 0 & 1 & 0 & 0 & 0 \\
1 & 0 & 0 & 0 & 0 & 0 & 0 & 0 \\
0 & 1 & 0 & 0 & 0 & 0 & 0 & 0 \\
0 & 0 & 1 & 0 & 0 & 0 & 0 & 0 \\
0 & 0 & 0 & 1 & 0 & 0 & 0 & 0 \\
0 & 0 & 0 & 0 & 0 & 1 & 0 & 0 \\
0 & 0 & 0 & 0 & 0 & 0 & 1 & 0 \\
0 & 0 & 0 & 0 & 0 & 0 & 0 & 1
\end{pmatrix}
\end{equation}
\begin{equation}
    (\mathcal{A}^5_{dec})^{\intercal} = 
\begin{pmatrix}
0 & 1 & 0 & 0 & 0 & 0 & 0 & 0 \\
0 & 0 & 1 & 0 & 0 & 0 & 0 & 0 \\
0 & 0 & 0 & 1 & 0 & 0 & 0 & 0 \\
0 & 0 & 0 & 0 & 1 & 0 & 0 & 0 \\
1 & 0 & 0 & 0 & 0 & 0 & 0 & 0 \\
0 & 0 & 0 & 0 & 0 & 1 & 0 & 0 \\
0 & 0 & 0 & 0 & 0 & 0 & 1 & 0 \\
0 & 0 & 0 & 0 & 0 & 0 & 0 & 1
\end{pmatrix}
\end{equation}
This procedure induces three extra vertices to the graph, as shown in Fig. \ref{5-cycle_graph_b}. 

Next, according to Eq. \eqref{k_shift_op} we write the shift operator for this system as 

\begin{equation}
S_{5C}= 
\begin{pmatrix}
(\mathcal{A}^5_{inc})^{\intercal} & 0 \\
0 & (\mathcal{A}^5_{dec})^{\intercal}
\end{pmatrix}
\end{equation}
The quantum circuit associated to this operator is shown in Fig. \ref{5-cycle_graph_c}.

%The minimum number of qubits necessary to build a quantum circuit that performs a DTQW on this topology is $n$ for the position register, plus an extra qubit for the coin register. 
The position register must be initialized in a composite state whose bitstring representation corresponds to the label (an integer number) of any vertex of the subgraph that conforms a $5$-cycle graph. If the composite state is initialized in any of the other vertices the walker will remain on that vertex every time the evolution operator is applied. The coin operator can be initialized either with the state $|0 \rangle$, $|1 \rangle$ or a superposition of both states. 

The action of the shift operator $S_{5C}$ on the quantum states of the position register associated to the labels of the vertices is presented in Table(\ref{table-shift-opS8}). From this example, notice that the minimum number of qubits to implement a DTQW on a $k$-cycle graph in the circuit model is $n+1$, where $n$ is the minimum value such that $2^n > k$.

\begin{table}[ht]
\begin{center}
\begin{minipage}{\textwidth}
\caption{Action of the shift operator for a DTQW on a 5-cycle (Fig. \ref{5-cycle_graph}) on the states $|0\rangle|q_1q_2q_3\rangle$ and $|1\rangle|q_1q_2q_3\rangle$. The first and second pair of columns represent the transition of position states with coin states $|0\rangle$ and $|1\rangle$, respectively} % title of Table
\centering % used for centering table
\begin{tabular*}{230pt}{@{\extracolsep{\fill}} c c c c @{\extracolsep{\fill}}} % centered columns (4 columns)
\toprule%,
\multicolumn{2}{c}{coin $|0\rangle$} & \multicolumn{2}{c}{coin $|1\rangle$} \\\cmidrule{1-2}\cmidrule{3-4}%
base 2 & base 10 & base 2 & base 10\\
\midrule
$|000 \rangle$ $\rightarrow$  $|001 \rangle$ & $|0 \rangle$ $\rightarrow$  $|1 \rangle$ & $|000 \rangle$ $\rightarrow$ $|100 \rangle$ & $|0 \rangle$ $\rightarrow$  $|4 \rangle$ \\
$|001 \rangle$ $\rightarrow$  $|010 \rangle$ & $|1 \rangle$ $\rightarrow$  $|2 \rangle$ & $|100 \rangle$ $\rightarrow$ $|011 \rangle$ & $|4 \rangle$ $\rightarrow$  $|3 \rangle$ \\
$|010 \rangle$ $\rightarrow$  $|011 \rangle$ & $|2 \rangle$ $\rightarrow$  $|3 \rangle$ & $|011 \rangle$ $\rightarrow$ $|010 \rangle$ & $|3 \rangle$ $\rightarrow$  $|2 \rangle$ \\
$|011 \rangle$ $\rightarrow$  $|100 \rangle$ & $|3 \rangle$ $\rightarrow$  $|4 \rangle$ & $|010 \rangle$ $\rightarrow$ $|001 \rangle$ & $|2 \rangle$ $\rightarrow$  $|1 \rangle$ \\
$|100 \rangle$ $\rightarrow$  $|000 \rangle$ & $|4 \rangle$ $\rightarrow$  $|0 \rangle$ & $|001 \rangle$ $\rightarrow$ $|000 \rangle$ & $|1 \rangle$ $\rightarrow$  $|0 \rangle$ \\
$|101 \rangle$ $\leftrightarrow$ $|101 \rangle$ & $|5 \rangle$ $\leftrightarrow$ $|5 \rangle$ & $|101 \rangle$ $\leftrightarrow$  $|101 \rangle$ & $|5 \rangle$ $\leftrightarrow$ $|5 \rangle$ \\
$|110 \rangle$ $\leftrightarrow$ $|110 \rangle$ & $|6 \rangle$ $\leftrightarrow$ $|6 \rangle$ & $|110 \rangle$ $\leftrightarrow$  $|110 \rangle$ & $|6 \rangle$ $\leftrightarrow$ $|6 \rangle$ \\
$|111 \rangle$ $\leftrightarrow$ $|111 \rangle$ & $|7 \rangle$ $\leftrightarrow$ $|7 \rangle$ & $|111 \rangle$ $\leftrightarrow$  $|111 \rangle$ & $|7 \rangle$ $\leftrightarrow$  $|7 \rangle$ \\  % [1ex]
%\botrule
\end{tabular*}
\label{table-shift-opS8} % is used to refer this table in the text
\end{minipage}
\end{center}
\end{table}

Notice that the set of adjacent transposition operators $\mathcal{T}_i$ has applications way beyond the cycle graph, given that it is a well known results in group theory that a general permutation is obtained by the product of transpositions \cite{rotman_1995} -- where a transposition is the permutation of two elements leaving the rest intact -- and any transposition can be obtained by the product of adjacent transpositions \cite{DLMF}. Thus, as we have provided the circuit form of all adjacent transpositions of rows in $2^n \times 2^n$ matrices, then any $2^n \times 2^n$ permutation matrix (or shunt in this work) can be generated by the set of operators $\mathcal{T}_i$. Combining this result with Eq. \eqref{additive_decomposition_eq}, we can generate the quantum circuit of associated to any transposed adjacency matrix.

\subsection{DTQW on the line}

\begin{figure}[h!]
\centering
%\captionsetup{justification=centering}
\includegraphics[scale=1]{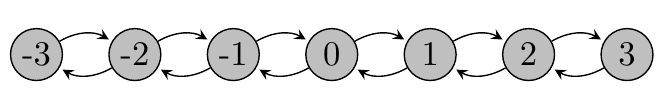}
\caption{7-line graph}
\label{line_graph_fig}
\end{figure}

Consider a graph constituted by $n$ vertices each one connected only to the previous and next ones, except for the vertices at the endings, which are only connected to the previous node. We refer to this mathematical object as the $n$-node line graph, or simply the $n$-line. Fig. \ref{line_graph_fig} displays a $7-line$ as an example. The adjacency matrix of an $n$-line is an $n \times n$ matrix presented in Eq. \eqref{n-line_adj_matrix}.
%, and for the purposes of DTQWs, we label its nodes with all the integers of the interval $[-n',n']$ in ascending order, where $n'=n/2-1$ . 

\begin{equation}
\mathcal{A}_{line} =
\label{n-line_adj_matrix}
\begin{pmatrix}
0 & 1 & 0 & 0 & \dots & 0 & 0 & 0 & 0 \\
1 & 0 & 1 & 0 & \dots & 0 & 0 & 0 & 0 \\
0 & 1 & 0 & 1 & \dots & 0 & 0 & 0 & 0 \\
0 & 0 & 1 & 0 & \dots & 0 & 0 & 0 & 0 \\
\vdots & \vdots & \vdots & \vdots & \ddots & \vdots & \vdots & \vdots & \vdots \\
0 & 0 & 0 & 0 & \dots & 0 & 1 & 0 & 0 \\
0 & 0 & 0 & 0 & \dots & 1 & 0 & 1 & 0 \\
0 & 0 & 0 & 0 & \dots & 0 & 1 & 0 & 1 \\
0 & 0 & 0 & 0 & \dots & 0 & 0 & 1 & 0 \\
\end{pmatrix}
\end{equation}
Notice that its form is quite similar to the adjacency matrix of a cycle graph, with the only difference that it only has elements in the diagonals above and below the main one, and not in any other entry. In order to implement a DTQW on an $n$-line, we can take advantage of this fact, and of what we derived in the previous subsection. 

\begin{figure}[t!]
    \centering
\subfigure[]{
\includegraphics[scale=0.65]{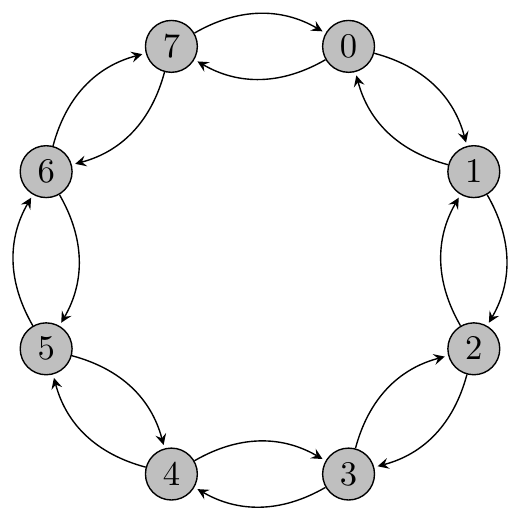}
\label{cycle_to_line_graph_fig_a}
}
\subfigure[]{
\includegraphics[scale=0.65]{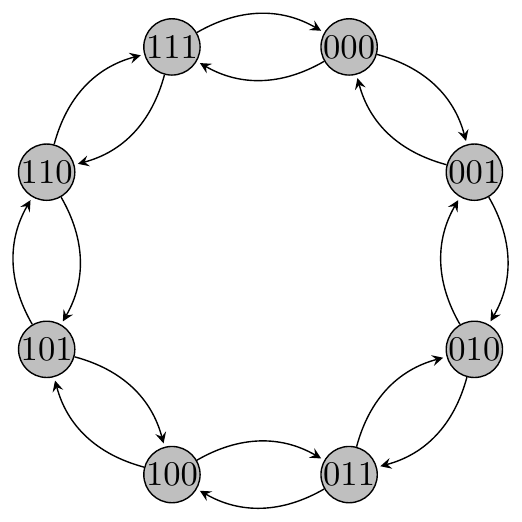}
\label{cycle_to_line_graph_fig_b}
}
\subfigure[]{
\includegraphics[scale=0.65]{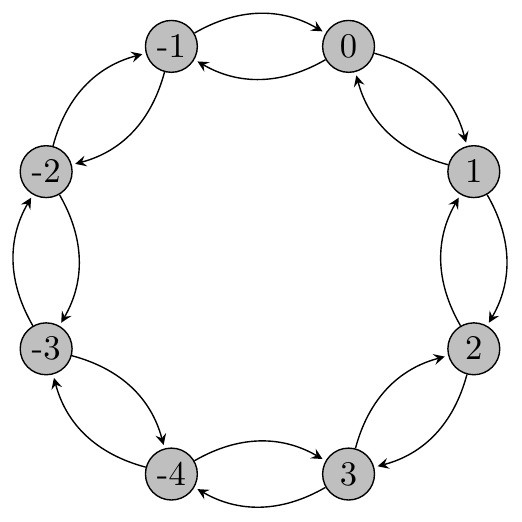}
\label{cycle_to_line_graph_fig_c}
}
\subfigure[]{
\includegraphics[scale=0.65]{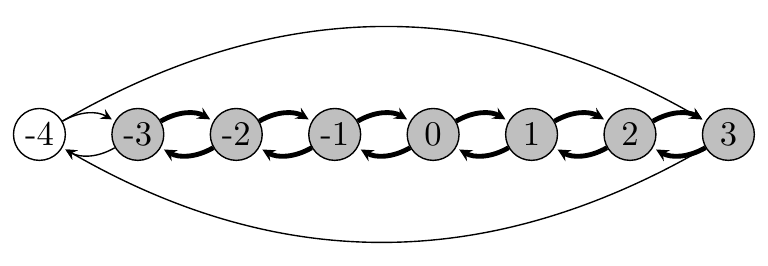}
\label{cycle_to_line_graph_fig_d}
}

\caption{Steps to transform a cycle graph into a line graph according to Table \ref{twos_complement_transformation_table}. (a) 8-cycle graph with labels in ascending order. (b) 8-cycle graph with transformed labels to binary. (c) 8-cycle graph with labels interpreted as two's complement (d) Rearranged 8-cycle graph into a straight line}
\label{cycle_to_line_graph_fig}
\end{figure}

We consider a $2^n$-cycle graph with nodes labeled with consecutive integers in ascending order starting from zero. Next we map the integer representation of each label to binary. After that, we consider the obtained bitstrings to be in two's complement notation\footnotemark[1], so that if we go back to integer representation, the upper half of the labels now take the values from $-1$ to $-2^n$.
\footnotetext[1]{Two's complement is a way to represent both positive and negative numbers in bitstring notation. Strings whose leftmost bit $b_n = 0$, correspond to positive numbers. When $b_n = 1$, we have a negative number. The rest of the bits, $N=b_{n-1}\dots b_{2}b_{1}$, represent the value of the number and $b_{n}$ only indicates the symbol. When $b_{n} = 0$, the right substring, $N$, has the same relation with decimal numbers as in binary notation. However, when $b_{n} = 1$, the relation of $N$ with decimal numbers is \textit{inverted} with respect to binary notation. E.g., for a 3-bit string, let $b_n = 1$, then for $N$ equals $00$, $01$, $10$ and $11$ the associated numbers are $4$, $3$, $2$ and $1$, respectively, in such a way that $100$, $101$, $110$ and $111$ are associated to $-4$, $-3$, $-2$ and $-1$, respectively.}
The proposed transformation for the numbers 0 to 8 is shown in Table \ref{twos_complement_transformation_table}. Finally, we rearrange the nodes of the graph, in ascending order in a straight line, and consider only the subgraph formed by the vertices with labels in the interval $[-n',n']$, where $n'=2^{n-1}-1$. Notice that $n'$ is always odd, thus there is always a central node in the line graphs studied in this work. An example of the process of transformation from an $8$-cycle graph to a $7$-line is shown in Fig. \ref{cycle_to_line_graph_fig}.

\begin{table}[ht]
\caption{Transformation of labels of an $8-cycle$ graph. Visual sequence given in Fig. \ref{cycle_to_line_graph_fig}} % title of Table
\label{twos_complement_transformation_table} 
\centering % used for centering table
\begin{tabular}{@{}c c c c c c c@{}} % centered columns (4 columns)
\toprule
Decimal & & Binary & & Two's complement & & Decimal \\ % inserts table
%heading
\midrule
0 & $\rightarrow$ & 000 & $\rightarrow$ & 000 & $\rightarrow$ & 0\\ 
1 & $\rightarrow$ & 001 & $\rightarrow$ & 001 & $\rightarrow$ & 1\\
2 & $\rightarrow$ & 010 & $\rightarrow$ & 010 & $\rightarrow$ & 2\\
3 & $\rightarrow$ & 011 & $\rightarrow$ & 011 & $\rightarrow$ & 3\\
4 & $\rightarrow$ & 100 & $\rightarrow$ & 100 & $\rightarrow$ & -4\\
5 & $\rightarrow$ & 101 & $\rightarrow$ & 101 & $\rightarrow$ & -3\\
6 & $\rightarrow$ & 110 & $\rightarrow$ & 110 & $\rightarrow$ & -2\\
7 & $\rightarrow$ & 111 & $\rightarrow$ & 111 & $\rightarrow$ & -1\\% [1ex] adds vertical space
%\botrule
\end{tabular}
\end{table}

In view of the transformation previously proposed, a DTQW on a line can be understood as a DTQW on a $2^n$-cycle graph whose labels have been mapped using two's complement and which holds the restriction that the evolution operator can be applied a maximum number of times $k \leq n'-n_0$, where $n' = 2^{n-1}-1$ and $n_0$ is the initial node of the walk. If it is applied a greater number of times than $k$, the walker will move in a cycle. 

This implies that instead of finding an additive decomposition for the adjacency matrix shown in Eq. \eqref{n-line_adj_matrix} and getting a new shift operator out of it, we can consider the shift operator already presented in Eq. \eqref{shift_2n-cycle}, and for the circuit implementation, we consider the circuit presented in Fig. \ref{inc-dec_operator}. The minimum number of qubits needed to implement a walk on a line of $k$ steps is obtained by solving Eq. \eqref{num_qubits_k_steps} for $n$ and then finding $\lceil n \rceil$+1, where the extra unit is due to the coin register. 

\begin{equation}
    \label{num_qubits_k_steps}
    k = 2^{n-1}-n_0-1
\end{equation}

\subsection{DTQW on the hypercube}

\begin{figure}[!hbt]
\centering
\includegraphics[]{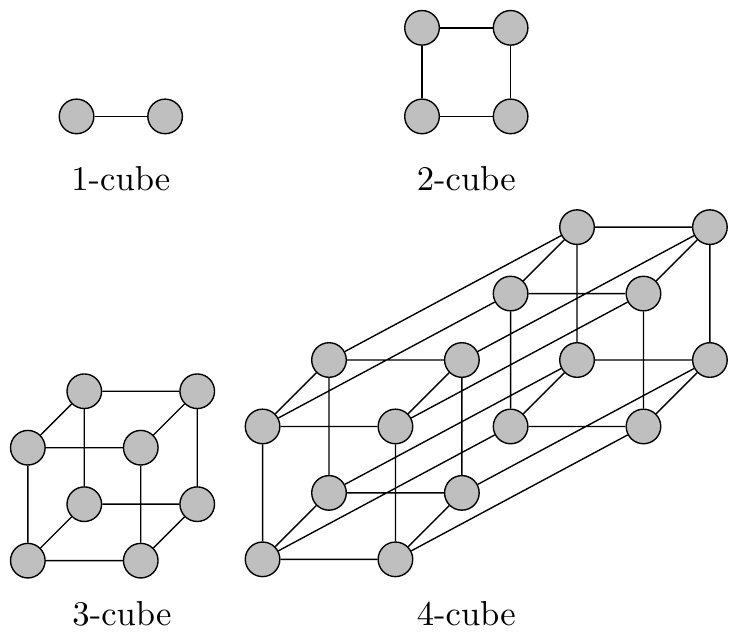}
\caption{Visual representation of the $n$-cube for $n$ = 1, 2, 3, 4}
\label{hypD0-D4}
\end{figure}

The hypercube graph, $Q_n(V,E)$, is an $n$-regular graph, built from the set of vertices and edges of the $n$-dimensional hypercube, also called $n-$cube. The graphical representation of $Q_n$ for n = 1, 2, 3, 4 is shown in Fig. (\ref{hypD0-D4}).

In order to construct the general graph $Q_n$, let $S_{n}$ be the set of all $n$-bitstings, and let $V=\{v_i: v_i \in S_{n}\}$ be the set of bitstrings that are labels for the vertices of the $n$-cube. Two such bitstrings $v_i$ and $v_j$ correspond to adjacent vertices if and only if they differ only in one bit, that is, their Hamming distance is equal to one. Adjacent vertices are connected by an edge $E_i \in E$, such that $E=\{\{v_i,v_j\} | v_i,v_j \in S_n; v_i \neq v_j\}$ is the set of edges of the graph. Each vertex $v_j$ is connected to $n$ other vertices by $n$ edges $E_i$. Notice according to section \ref{section2} a DTQW takes place on a directed graph, thus, to comply with this requirement, we can replace edges $E_i$ by a set of parallel arcs pointing in opposite directions.
 
The usual definition of the shift operator for a DTQW on an $n$-cube is given in Eq. \eqref{common_shift_op_n-cube}.

\begin{equation} 
    S=\sum_{i=0}^{m-1}\sum_{j=0}^{2^m-1} |c_i,v_j \oplus e_i \rangle \langle c_i, v_j| 
    \label{common_shift_op_n-cube}
\end{equation}
or
\begin{align*}
    S=\sum_{i=0}^{m-1}\sum_{j=0}^{2^m-1} |c_i\rangle \langle c_i|\otimes |v_j \oplus e_i \rangle \langle v_j|\hspace{-1.7em}
\end{align*}
where $e_i$ is the $n$-dimensional bitstrings with all bits zero except for the $i$th bit, which is 1, $c_i$ and $v_j$ are $m$ and $n$-dimensional bitstring, respectively, and the symbol $\oplus$ represents the binary sum or bit-wise xor. Notice that in this system the dimension of the coin spacace, $m$, is related to the dimension of the position space, $n$, as $n=2^m$. 

The factor $|c_i\rangle \langle c_i|$ restricts $S$ to be a block diagonal matrix, and thus the block diagonal elements, $\sum_{j=0}^{2^m-1} |v_j \oplus e_i \rangle \langle v_j|$ (for fixed $i$), are the shunt decomposition of the adjacency matrix of a $2^m$-cube.

The adjacency matrix of an $n$-cube can be obtained following the next recursive pattern presented in \cite{florkowski_hypcube_ad_mat}

\begin{equation}
\mathcal{A}_{n+1} =
\label{general_hypercube_matrix}
\begin{pmatrix}
\mathcal{A}_n & I_n \\
I_n & \mathcal{A}_n
\end{pmatrix}
;\; n > 0 ;\; \mathcal{A}_0 = (0)
\end{equation}
For any dimension $n=2^m$, $\mathcal{A}_n$ can be decomposed as the sum of unitary matrices, $\mathcal{A}_{2^m} = \sum_{i=0}^{n-1} \mathcal{P}^{\intercal}_i$, where  each $\mathcal{P}^{\intercal}_i$ is written in terms of the Pauli matrix $\sigma_x$ and the $2 \times 2$ identity only

\begin{figure}[t!]
\centering
\subfigure[]{
\includegraphics[scale=0.8]{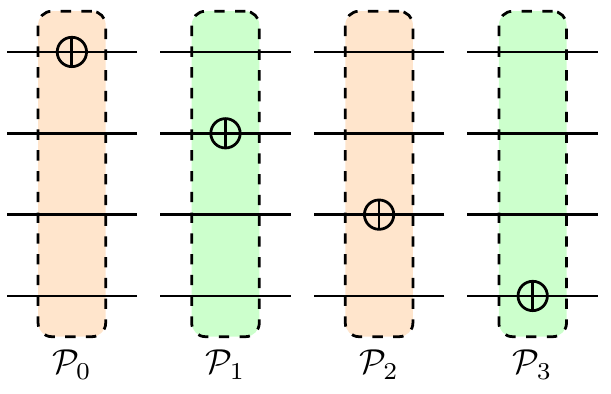}
\label{4q-cnot-network_a}
}
\subfigure[]{
\includegraphics[scale=0.8]{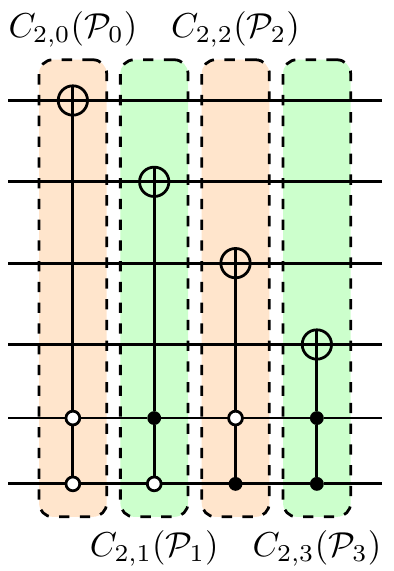}
\label{4q-cnot-network_b}
}
\subfigure[]{
\includegraphics[scale=0.8]{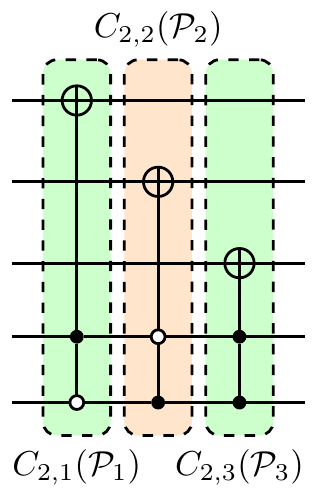}
\label{4q-cnot-network_c}
}

\caption{(a) Quantum gate representation of the matrices that conform the shunt decomposition of the adjacency matrix of a 4-cube. Quantum circuit associated to the shift operator of a DTQW on a (b) 4-cube and (c) 3-cube with self-loops}
\label{4q-cnot-network}
\end{figure}

%For any dimension $n=2^m$, $\mathcal{A}_n$ can be decomposed into a sum of unitary matrices which take the following form

\begin{equation}
    \label{summands_n-cube}
    \mathcal{P}^{\intercal}_i = I_2^{\otimes i} \otimes \sigma_x \otimes I_2^{\otimes (n-i-1)} ; \; i= 0, 2, \dots ,n-1
\end{equation}
where we define $I_2^{\otimes 0} = (1)$.

For the case in which $v_j$ is the bitstring whose binary value corresponds to decimal value of the index $j$, then the following equation holds for some bitstring $e_i$

\begin{equation}
    \mathcal{P}^{\intercal}_i = I_2^{\otimes i} \otimes \sigma_x \otimes I_2^{\otimes (n-i-1)}=\sum_{j=0}^{2^m-1} |v_j \oplus e_i \rangle \langle v_j|
\end{equation}
This can be corroborated by explicit matrix analysis of both expressions.

Therefore, according to Eq. \eqref{block_diag_shift}, the matrix representation of the shift operator is given by
\begin{equation}
    \label{shift_n-cube_eq}
    S = \bigoplus\limits_{i=0}^{n-1}I_2^{\otimes i} \otimes \sigma_x \otimes I_2^{\otimes (n-i-1)}
\end{equation}

In order to get the quantum circuit associated with $S$, we use the fact that a quantum gate can be built for every unitary matrix $\mathcal{P}^{\intercal}_i$ by setting a register of $n$ qubits, and placing a $\sigma_x$ gate in the $i$th qubit leaving the rest of the register empty, as in the example shown in Fig. \ref{4q-cnot-network_a}. Next, we control all gates $\mathcal{P}^{\intercal}_i$ using all the qubits of the coin register with a different pattern of black and white controls for each $\mathcal{P}^{\intercal}_i$, in order to obtain the quantum gates $C_{m,j}(\mathcal{P}^{\intercal}_i)$, as explained in section \ref{section3}. The combination of gates $C_{m,j}(\mathcal{P}^{\intercal}_i)$ is equal to the shift operator of the system, and form a pattern as in the example shown in Fig. \ref{4q-cnot-network_b}.

Furthermore, we can use the circuit of a $2^m$-cube to obtain a circuit for any cube of dimension lower than $2^m$, with the restriction that the number of dimension we decrease from the original cube, will be the number of self-loops added to all the vertices of the new structure. To do this, simply remove the first $k$ CNOTs with the largest range; the resulting circuit will be that of a $(2^m-k)$-cube with $k$ self-loops. For instance, if we want to obtain a circuit for a $3$-cube with self-loops, then we modify the circuit of a $4$-cube to obtain the circuit shown in Fig. \ref{4q-cnot-network_c}.

\begin{figure}[t!]
\centering
\subfigure[]{
\includegraphics[scale=0.5]{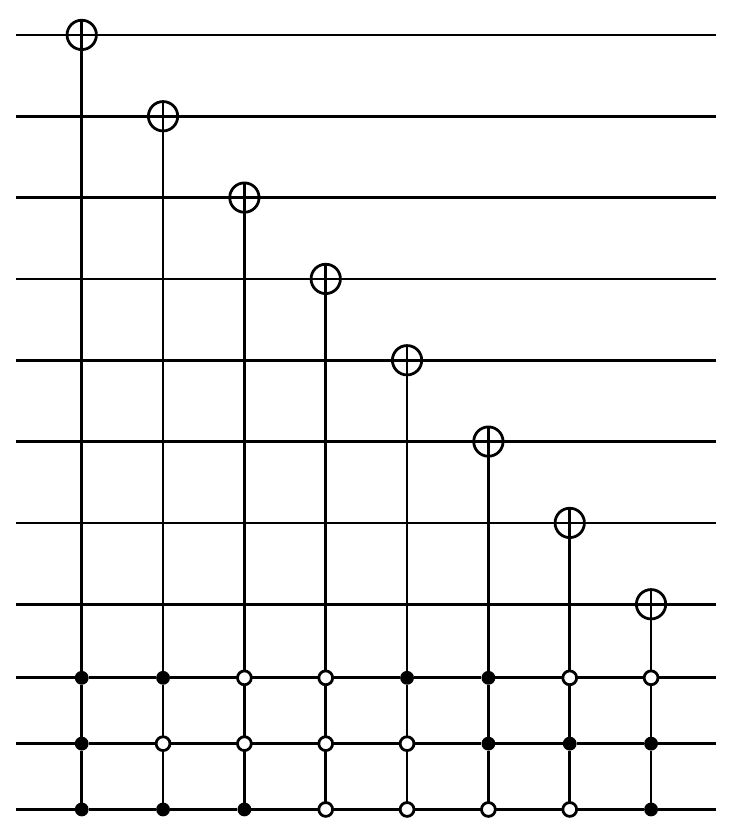}
\label{circ-hyp-8d_a}
}
\subfigure[]{
\includegraphics[scale=0.5]{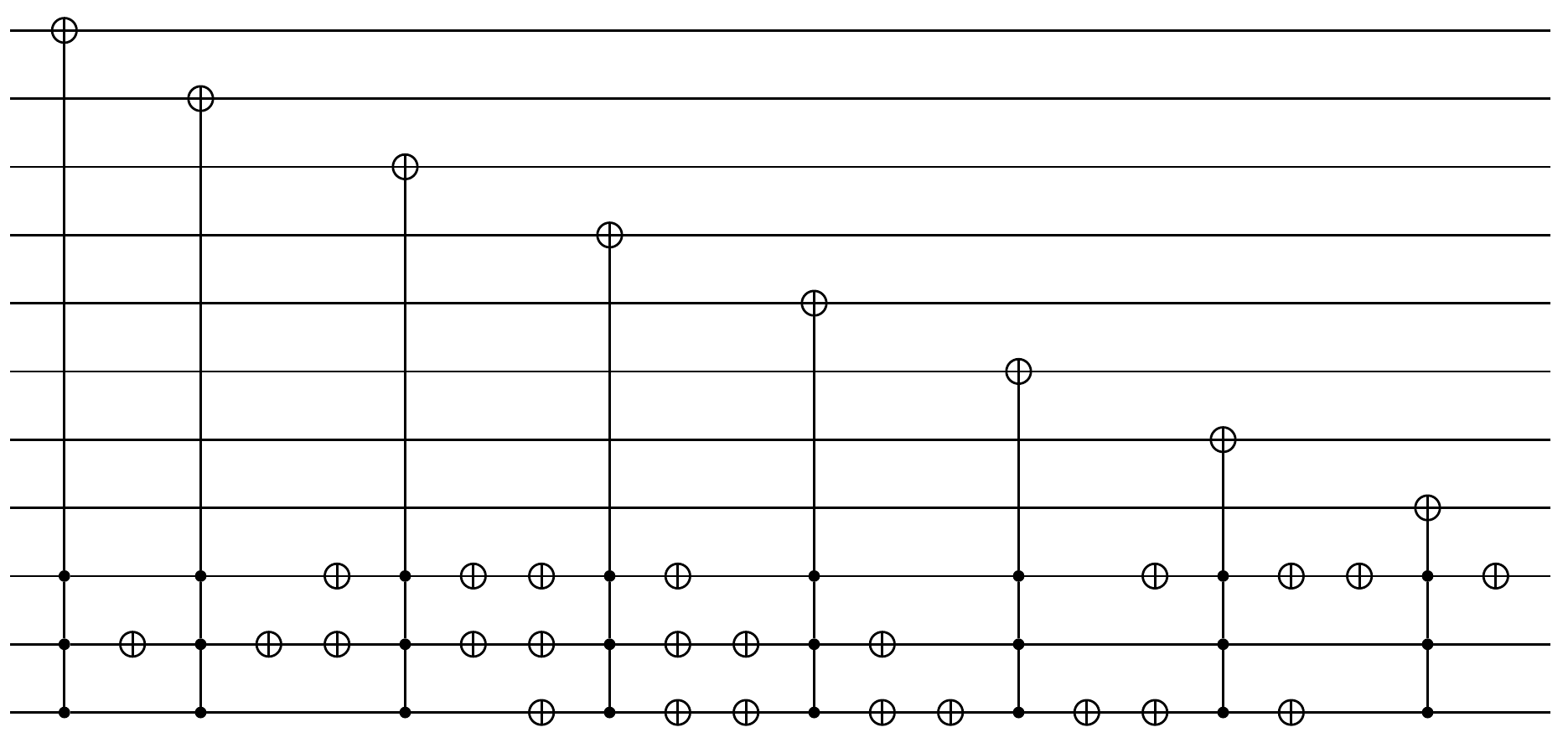}
\label{circ-hyp-8d_b}
}

\subfigure[]{
\includegraphics[scale=0.5]{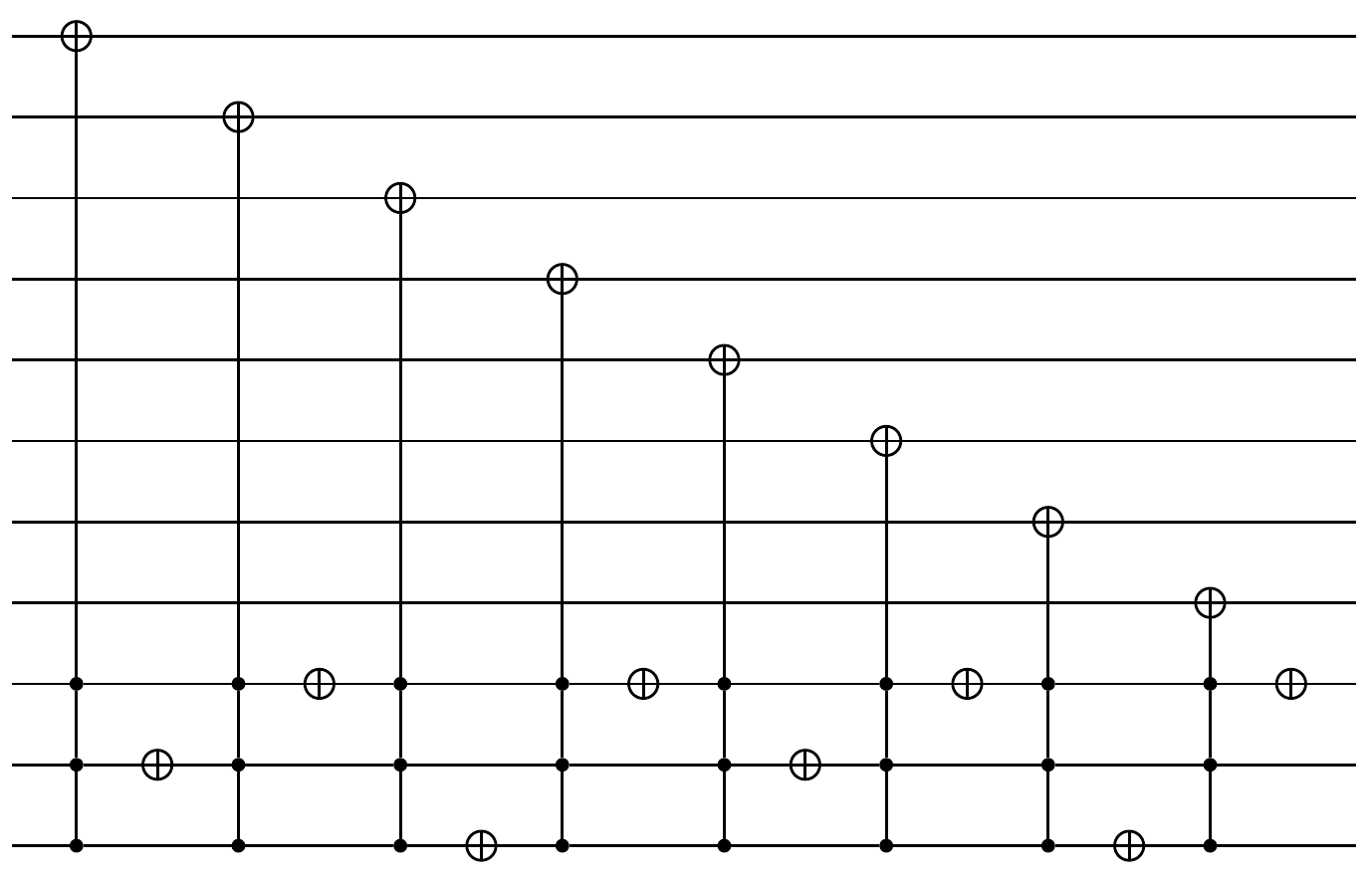}
\label{circ-hyp-8d_c}
}
\caption{(a) Circuit implementation of the shift operator of a DTQW on a 8-cube proposed in \cite{douglas_wang_2009}, where the controls follow gray code sequence. (b) Current quantum computers only admit black controls, thus white controls must replace by black controls surrounded by two $\sigma_x$ gates. (c) Redundant CNOTs are cancelled out, which could represent an advantage if the circuit is to be implemented in noisy quantum computers}
\label{circ-hyp-8d}
\end{figure}

\raggedbottom
The general pattern of the quantum circuit representation of $S$ was proposed by Douglas and Wang in \cite{douglas_wang_2009}, being a key element to our paper to construct an explicit link  between the shunt decomposition method and the quantum circuit representation of $S$. In order to simplify the circuit, notice that negated black controls are represented by white controls, and we negate a black control by placing $\sigma_x$ gates at both sides of it. Current quantum computers do not allow the direct implementation of white controls, thus if we are to build a quantum circuit that consists of a sequence of multi-control gates, we want that when we replace all white controls with black controls sorrounded by $\sigma_x$ gates to the left and to the right, as many $\sigma_x$ gates as possible cancel out. This can be achieved by initially designing the circuit in such a way that as many white controls are next to each other in the sequence of multi-control gates. For this purpose, we propose the use of gray code, which a bitstring representation of numbers such that for two consecutive numbers in decimal notation the corresponding bitstrings differ in only one bit. That is, this sequence maximizes the resemblance of consecutive bitstrings, and if we associate 0's and 1's with white and black controls, respectively, when using this sequence to generate all $C_{m,j}(\mathcal{P}^{\intercal}_i)$ gates that constitute the shift operator of a $2^n$-cube, we can cancel out a wide number of redundant $\sigma_x$ gates. Although, for gray code to be efficient, we must start the sequence with the string of all ones, or to put it in another way, the leftmost (or rightmost, it is indistinct) gate must be controlled with only black controls, and then single-bit-flipping sequence can be used to control the following gates. This is exemplified in Fig. \ref{circ-hyp-8d}.

%an optimal form to arrange all $C^{m}_j(\mathcal{P}^{\intercal}_i)$ gates is shown in Fig. \ref{circ-hyp-8d_b}, given that this arrangement allows cancellation of $\sigma_x$ gates.

\subsection{DTQW on the $2^m$-complete graph with self-loops}

\begin{figure}[h!]
\centering
\includegraphics[scale=0.7]{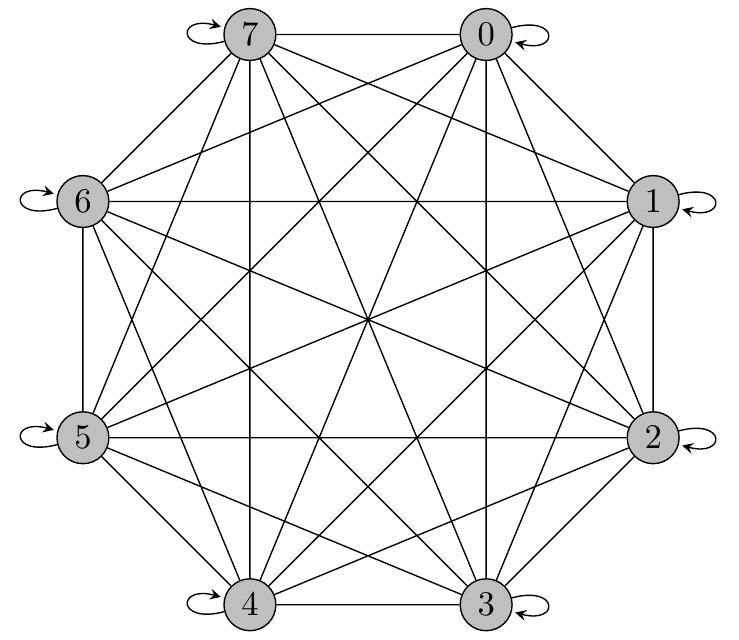}
\caption{Graph $\mathcal{K}_8$}
\label{k8_graph_self_loops_fig}
\end{figure}

An $n$-complete graph with self-loops, $\mathcal{K}_n$ is a set of $n$ vertices such that the vertex $v_i$ is connected with all other vertices $v_j$, including itself. An example of this type of graph is given in Fig. (\ref{k8_graph_self_loops_fig}).

The adjacency matrix of an $n-$node complete graph with self-loops at every vertex is a matrix whose entries are all ones (Eq. \eqref{adj_mat_n-complete_graph}). Here we propose a method to decompose the adjacency matrix of the $\mathcal{K}_{2^m}$ graph, where $m$ is the dimension of the coin register.

\begin{equation}
\label{adj_mat_n-complete_graph}
\mathcal{A}_{\mathcal{K}_n}=
    \begin{pmatrix}
1 & 1 & 1 & \dots & 1 & 1 & 1 \\
1 & 1 & 1 & \dots & 1 & 1 & 1 \\
1 & 1 & 1 & \dots & 1 & 1 & 1 \\
\vdots & \vdots & \vdots & \ddots & \vdots & \vdots & \vdots \\
1 & 1 & 1 & \dots & 1 & 1 & 1 \\
1 & 1 & 1 & \dots & 1 & 1 & 1 \\
1 & 1 & 1 & \dots & 1 & 1 & 1 
\end{pmatrix}
\end{equation}

The decomposition can be done by means of the tensor products of $\sigma_x$ and $I_2$ operators. For this purpose, we appeal again to binary code: consider the $\sigma_x$ operator to be a $"1"$, and $I_2$ operator to be a $"0"$. Now consider the tensor product of a number $n_1$ of operators $\sigma_x$  and a number $n_2$ of operators $I_2$ ($n_1+n_2=m$), in such a way that their sequence is interpreted as binary code. Every sequence of tensor products defines a matrix $\mathcal{P}^{\intercal}_i$, such that the index $i$ corresponds to the number the sequence represents. The sequence of operators for the graph $\mathcal{K}_8$ is shown in the set of Eqs. \eqref{seq_Mi_K8}.

\begin{subequations}
\begin{align}
    \mathcal{P}^{\intercal}_0 = I_2 \otimes I_2 \otimes I_2 \\
    \mathcal{P}^{\intercal}_1 = I_2 \otimes I_2 \otimes \sigma_x \\
    \mathcal{P}^{\intercal}_2 = I_2 \otimes \sigma_x \otimes I_2 \\
    \mathcal{P}^{\intercal}_3 = I_2 \otimes \sigma_x \otimes \sigma_x \\
    \mathcal{P}^{\intercal}_4 = \sigma_x \otimes I_2 \otimes I_2 \\
    \mathcal{P}^{\intercal}_5 = \sigma_x \otimes I_2 \otimes \sigma_x \\
    \mathcal{P}^{\intercal}_6 = \sigma_x \otimes \sigma_x \otimes I_2 \\
    \mathcal{P}^{\intercal}_7 = \sigma_x \otimes \sigma_x \otimes \sigma_x
\end{align}
\label{seq_Mi_K8}
\end{subequations}
Then
\begin{equation}
\mathcal{A}_{\mathcal{K}_8}=
\sum_{i=0}^{7} \mathcal{P}^{\intercal}_i
\label{adjacency_M_8_decomposition-eq}
\end{equation}

\begin{figure}
\centering
\subfigure[]{
\includegraphics[scale=0.65]{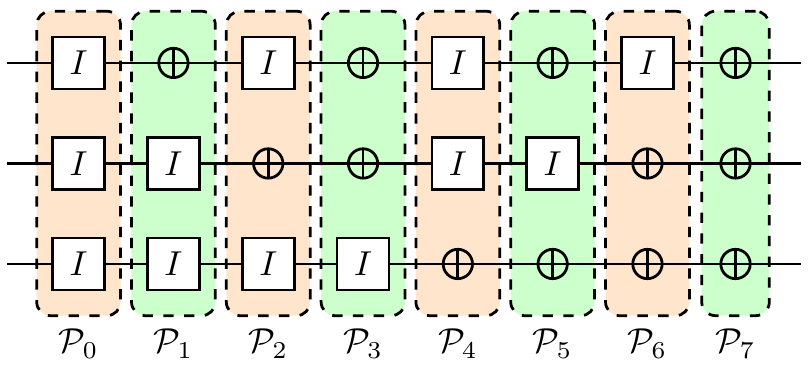}
\label{sequence_K8_circuit_a}
}
\subfigure[]{
\includegraphics[scale=0.65]{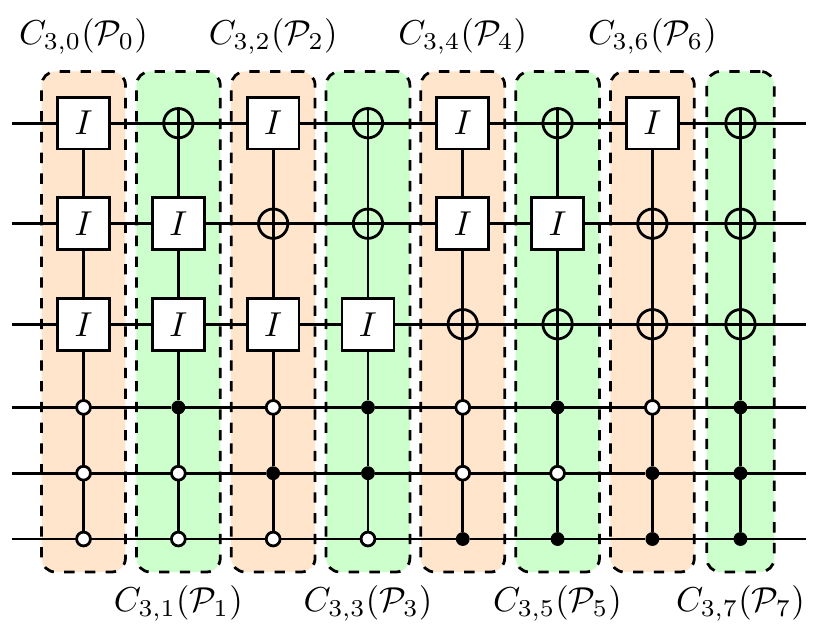}
\label{sequence_K8_circuit_b}
}
\caption{(a) Gate sequence corresponding to the permutation matrices presented in the set of Eqs. \eqref{seq_Mi_K8}. (b) Gate sequence corresponding to the shift operator of a DTQW on $\mathcal{K}_8$ (Fig. (\ref{k8_graph_self_loops_fig})). Notice that the sequence of identity and CNOT gates is the same as the sequence of black and white dots for each gate} 
\label{sequence_K8_circuit}
\end{figure}

Fig. \ref{sequence_K8_circuit_a} shows the sequence of quantum gates that corresponds to every operator $\mathcal{P}^{\intercal}_i$. Similar to the previous sections, we have to control these gates and then place them in sequence in order to obtain a quantum circuit for a shift operator that performs a DTQW on the graph $\mathcal{K}_8$. In this case, the quantum circuit can be simplified if the index $j$ of the controlled gate, $C_{m, j}(\mathcal{P}^{\intercal}_i)$, is equal to the index of the operator $\mathcal{P}^{\intercal}_i$. This means that every sequence of $\sigma_x$ and $I_2$ gates will be controlled by a sequence of white and black dots that represent the same number in binary, as shown in Fig. \ref{sequence_K8_circuit_b}.

\begin{figure}[b!]
\begin{minipage}[a]{\textwidth}
\centering
\subfigure[]{
\includegraphics[scale=0.8]{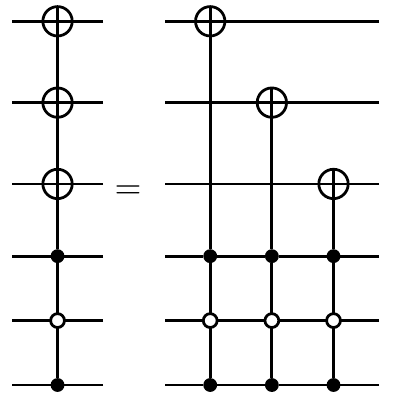}
\label{multi-control_simplification_a}
}
\subfigure[]{
\includegraphics[scale=0.8]{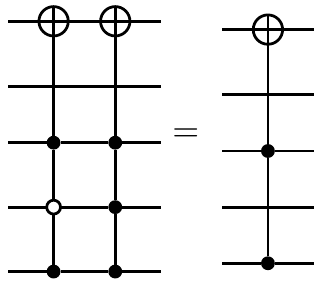}
\label{multi-control_simplification_b}
}
\caption{\label{multi-control_simplification} Examples of the (a) Decomposition of a multi-control multi-target gate. (b) Synthesis of two multi-control single-target gates which differ by only one control color for the same control qubit}
\end{minipage}
\end{figure}

The addition of the identity gate to the circuits is just for illustrative purposes, it can be neglected. Although, even if we do so, some of the gates $C_{m, j}(\mathcal{P}^{\intercal}_i)$ are still multi-target multi-control. This type of gate can be decomposed into a set of single-target multi-control gates \cite{daraeizadeh_kumar_2014}, as shown in the example of Fig. \ref{multi-control_simplification_a}. Then, multi-controlled gates that have the same target qubit and whose sequence of black and white dots differs only in one element can be simplified as shown in the example of Fig. \ref{multi-control_simplification_b}. This property has been derived in different works \cite{rahman_rice_2014, cheng_guan_wang_zhu_2012, arabzadeh_saeedi_zamani_2010} using different methods.

\begin{figure}[!tbp]
\centering
\begin{minipage}[t]{0.45\textwidth}
\centering
\subfigure[]{
\includegraphics[scale=0.7]{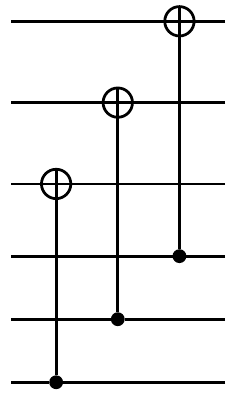}
\label{Kn_simplified_circuit_a}
}
\subfigure[]{
\includegraphics[scale=0.7]{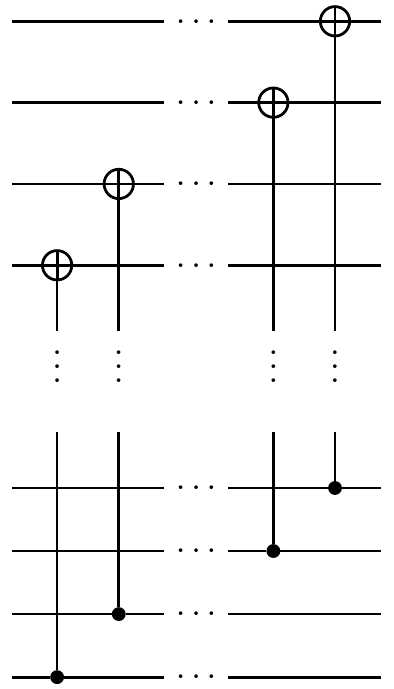}
\label{Kn_simplified_circuit_b}
}
\caption{(a) Synthesized circuit after applying the decompositions shown in Fig. \ref{multi-control_simplification} to the shift operator of a DTQW on $\mathcal{K}_8$ (Fig. \ref{sequence_K8_circuit_b}). (b) Shift operator composed of $n$ CNOT gates to perform a DTQW on $\mathcal{K}_{2^n}$}
\label{Kn_simplified_circuit}
  \end{minipage}
\hspace{.2in}
\begin{minipage}[t]{0.47\textwidth}
\centering
\subfigure[]{
\includegraphics[scale=0.6]{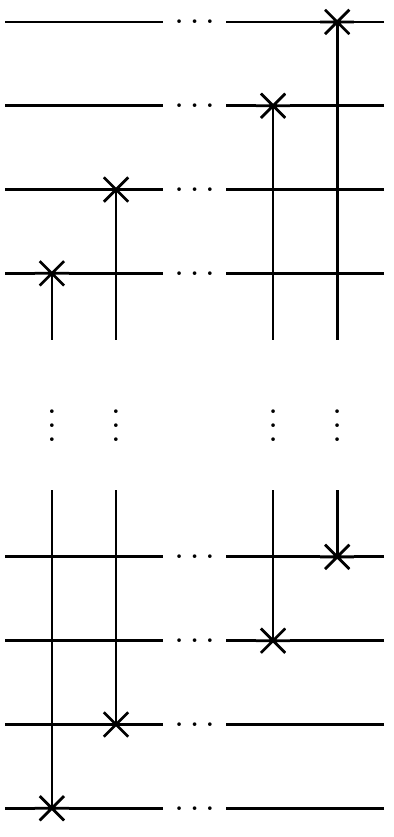}
\label{Wang_complete_graph_circuit_a}
}
\subfigure[]{
\includegraphics[scale=0.6]{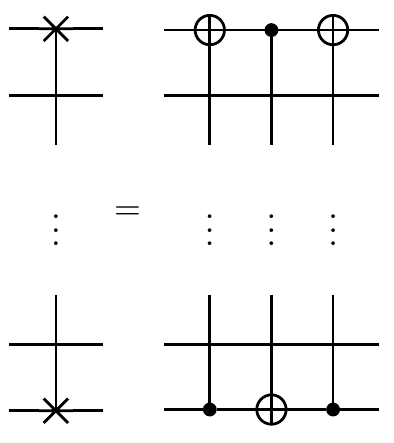}
\label{Wang_complete_graph_circuit_b}
}
\caption{(a) Sequence of $n$ SWAP gates corresponding to the shift operator of a DTQW on $\mathcal{K}_{2^n}$, proposed in \cite{douglas_wang_2009}. (b) Decomposition of a SWAP gate into CNOT gates}
\label{Wang_complete_graph_circuit}
\end{minipage}
\end{figure}

When both properties are applied to the whole network of multi-control CNOT gates shown in Fig. (\ref{multi-control_simplification}), everything can be reduced to only three gates, as shown in Fig. \ref{Kn_simplified_circuit_a}. The sequence of controlled gates can be continued (see Fig. \ref{Kn_simplified_circuit_b}) to obtain a shift operator for a DTQW on a complete graph with $2^m$ vertices. %The evolution operator is obtained by placing a Hadamard gate in every qubit of the coin register before the sequence of gates that conforms the shift operator.
Notice that we can use the same procedure replacing the gate $\sigma_x$ by an arbitrary multi-qubit $U$ gate to obtain efficient implementation of operators shift for different graphs.

In terms of outer products, the shift operator for this system can be defined in a similar way than the usual definition of the shift operator for a DTQW on the hypercube Eq. \eqref{common_shift_op_n-cube}, with the only difference that in this case, the direction bitstrings $e_i$ are all bitstrings of size $m$, in such a way that we associate one bitstring to each edge connected to the vertex where the quantum walker stands on. This provides $2^m$ directions for the walker to move at each vertex. The analysis of the matrix form of $S$ resulting from this idea conducted us to find the decomposition presented in the set of Eqs. \eqref{seq_Mi_K8}, which can be extended to any dimension $2^m$ following the same procedure.

Douglas and Wang proposed in \cite{douglas_wang_2009} a circuit which can be generalized to perform a DTQW on a graph $\mathcal{K}_{2^n}$, using $n$ SWAP gates, arranged as presented in Fig. \ref{Wang_complete_graph_circuit}. The circuit was the result of mapping a different shift operator to perform a DTQW on the same topology. The shift operator used has the following action on the state of a walker $S|c_i,v_j\rangle=|v_j,c_i\rangle$, thus it seems logical to use SWAP operators for its mapping to a quantum circuit. Each SWAP gate can be implemented using three CNOT gates, therefore the shift operator proposed by Douglas and Wang can be implemented using $3n$ CNOT gates, while the circuit we propose for a DTQW on the same topology, needs only $n$ CNOT gates, as can be seen from Fig. \ref{Kn_simplified_circuit_b}, which proves it to be more efficiently implementable in terms of CNOT gates.

\section{\label{section5}Simulation and Experimental Implementation of DTQWs}

\begin{figure}[h!]
\centering
\subfigure[]{
\includegraphics[scale=0.8]{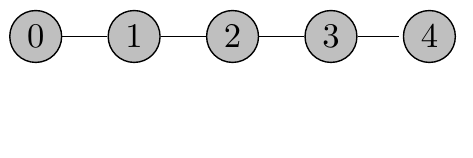}
\label{connectivity_map_a}
}
\subfigure[]{
\includegraphics[scale=0.8]{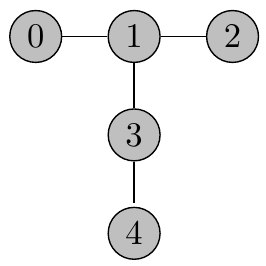}
\label{connectivity_map_b}
}
\caption{Connectivity map of the quantum processors used in this work to perform DTQWs. (a) Processor type: Falcon r5.11L. ibmq\_manila and ibmq\_bogota follow this connectivity. (b) Processor type: Falcon r4T. ibmq\_quito and ibmq\_lima follow this connectivity. See \cite{ibm_quantum_processors} for more information on the processors}
\label{connectivity_map}
\end{figure}

In this section we implement specific cases of the quantum circuits associated to the shift operators of the topologies studied in the previous sections, using both IBM's quantum computers publicly available through IBM Quantum \cite{ibm_quantum} and Qiskit Aer simulator. The topologies that we will focus on in this section are the graph $\mathcal{K}_4$, the 8-line, the 3-cube with self-loops and the 5-cycle graph, presented in this order in the coming subsections. For the first two graphs experimental results will be presented, while for the last two graphs we limit ourselves to present Qiskit simulations. The quantum processors we used to perform the experimental computations are: ibmq\textunderscore manila, ibmq\textunderscore bogota, ibmq\textunderscore quito, and ibmq\textunderscore lima, with quantum volume (QV) 32, 32, 16, and 8, respectively. All of these quantum processors have 5 qubits, connected in T-like (quito and lima) and line-like (manila and bogota) topologies as shown in Fig. (\ref{connectivity_map}). Due to the low amount of available qubits we restrict our DTQWs to take place in low-dimensional cases of the graphs studied. We implemented the quantum circuits directly in the Quantum Composer.

We utilized the statistical distance between distributions 
\begin{equation}
    \ell_1(P,Q) = \frac{1}{2}\sum\limits_{\forall i}|P(i)-Q(i)|
\end{equation}
as a metric to evaluate the performance of each quantum circuit compared to the theoretical distributions, where the shorter the distance, the higher the resemblance of an experimental distribution to the theoretical one. %We calculated the $\ell_1$ distance between the theoretical and experimental distributions of all the models and steps, and summarized that information in Tables \ref{5cycle_table} to \ref{K4_Wang_table}, in which shorter distances for each step are in bold font. 
To have a visual representation of the performance of the implemented DTQWs, we present a summary grid of the obtained probability distributions in the following sections, that include the theoretical distributions and the experimental distributions with best metrics for each model and each step, regardless of the quantum processor used. In all cases, i.e. simulated, experimental and theoretical DTQWs, the initial state of the walker was the state $\psi = |0\rangle_C \otimes |0\rangle_P$, for all the models. The \textit{theoretical} distributions were calculated analytically, using the corresponding shift and coin operators of each graph, obtained as described in the previous sections. 

%The original and transpiled one-step-circuits of the models with best metrics, the depth and size of all the circuits and the calibration parameters of the quantum computers used, can be found in appendices \ref{appx_quantum_circuits}, \ref{appx_depth_size}, and \ref{appx_calibrations}, respectively. 

A repository with all the circuits resulting from this work in Qiskit format, is available in \cite{github_repo}.

\subsection{DTQW on the graph $\mathcal{K}_4$}
\label{K4_section}
To perform a DTQW on a graph $\mathcal{K}_{2^m}$ we need $2m$ qubits, thus, with the available 5-qubit quantum computers we can only run circuits for the graphs $\mathcal{K}_2$ and $\mathcal{K}_4$, although we will focus on the latter. We compare the performance of both models for the graph $\mathcal{K}_4$ shown in Figs. \ref{Kn_simplified_circuit} and \ref{Wang_complete_graph_circuit}. The metrics of the experimental distributions are shown in Tables \ref{K4_Wing_table} and \ref{K4_Wang_table}, where the former corresponds to the model introduced in this paper ($CNOT$ model), while the latter corresponds to the model introduced by Douglas and Wang ($SWAP$ model) \cite{douglas_wang_2009}. The depth and size of the original single-step circuits is 2 and 4, respectively, for both models. However, to be able to run in IBM's quantum computers, these circuits must undergo a transpilation process to adapt to the topology (see Fig. \ref{connectivity_map}) of the quantum computers and be optimized for noise reduction. Furthermore, this process involves a change of basis to the set $CNOT$, $\sqrt{x}$ and $R_z$, which is the native set of gates in IBM's quantum computers. Thus, in the case of the one-step transpiled circuits, the size of the circuit with best metric is 11 and 14, for the $CNOT$ model, while the depth and size of the circuit with best metric for the $SWAP$ model is 15 and 18, respectively.

\begin{table}[h!]
\caption{$\ell_{1}$ distance between experimental and theoretical distributions for the graph $\mathcal{K}_4$ using the $CNOT$ model. ibmq\textunderscore quito has the best overall performance.
%Distances are short in general, which is an indicator of good experimental performance. This coincides with the fact that experimental probability distributions (Fig. \ref{k4_wing_all_distros}) resemble the theoretical ones.
} % title of Table
\label{K4_Wing_table} 
\centering % used for centering table
\begin{tabular}{@{}c c c c c@{}} % centered columns (4 columns)
\toprule
Device/Steps & 1 step & 2 steps & 3 steps & Avg\\ [0.5ex] % inserts table
%heading
\midrule
ibmq\textunderscore manila & 0.057 & 0.114 & \textbf{0.297}  & 0.156\\ 
ibmq\textunderscore bogota & 0.114 & 0.091 & 0.352 & 0.185\\
ibmq\textunderscore quito & 0.048 & \textbf{0.024} & 0.319 & \textbf{0.130}\\
ibmq\textunderscore lima & \textbf{0.040} & 0.049 & 0.384 & 0.158\\ [1ex] % [1ex] adds vertical space
%\botrule
\end{tabular}
\end{table}
\begin{table}[h!]
\caption{$\ell_{1}$ distance between experimental and theoretical distributions for the graph $\mathcal{K}_4$ using the $SWAP$ model. ibmq\textunderscore manila has the best overall performance.
%Distances are short in general, which is an indicator of good experimental performance. This coincides with the fact that experimental probability distributions (Fig. \ref{k4_wang_all_distros}) resemble the theoretical ones. 
} % title of Table
\label{K4_Wang_table} 
\centering % used for centering table
\begin{tabular}{@{}c c c c c@{}} % centered columns (4 columns)
\toprule
Device/Steps & 1 step & 2 steps & 3 steps & Avg\\ [0.5ex] % inserts table
%heading
\midrule
ibmq\textunderscore manila & 0.051 & 0.130 & \textbf{0.182} & \textbf{0.121}\\ 
ibmq\textunderscore bogota & 0.130 & 0.144 & 0.227 & 0.167\\
ibmq\textunderscore quito & 0.112 & 0.101 & 0.341 & 0.185\\
ibmq\textunderscore lima & \textbf{0.048} & \textbf{0.054} & 0.355 & 0.152\\ [1ex] % [1ex] adds vertical space
%\botrule
\end{tabular}
\end{table}

\begin{figure}[h!]
    \centering
    \subfigure[]{
    \includegraphics[scale=0.53]{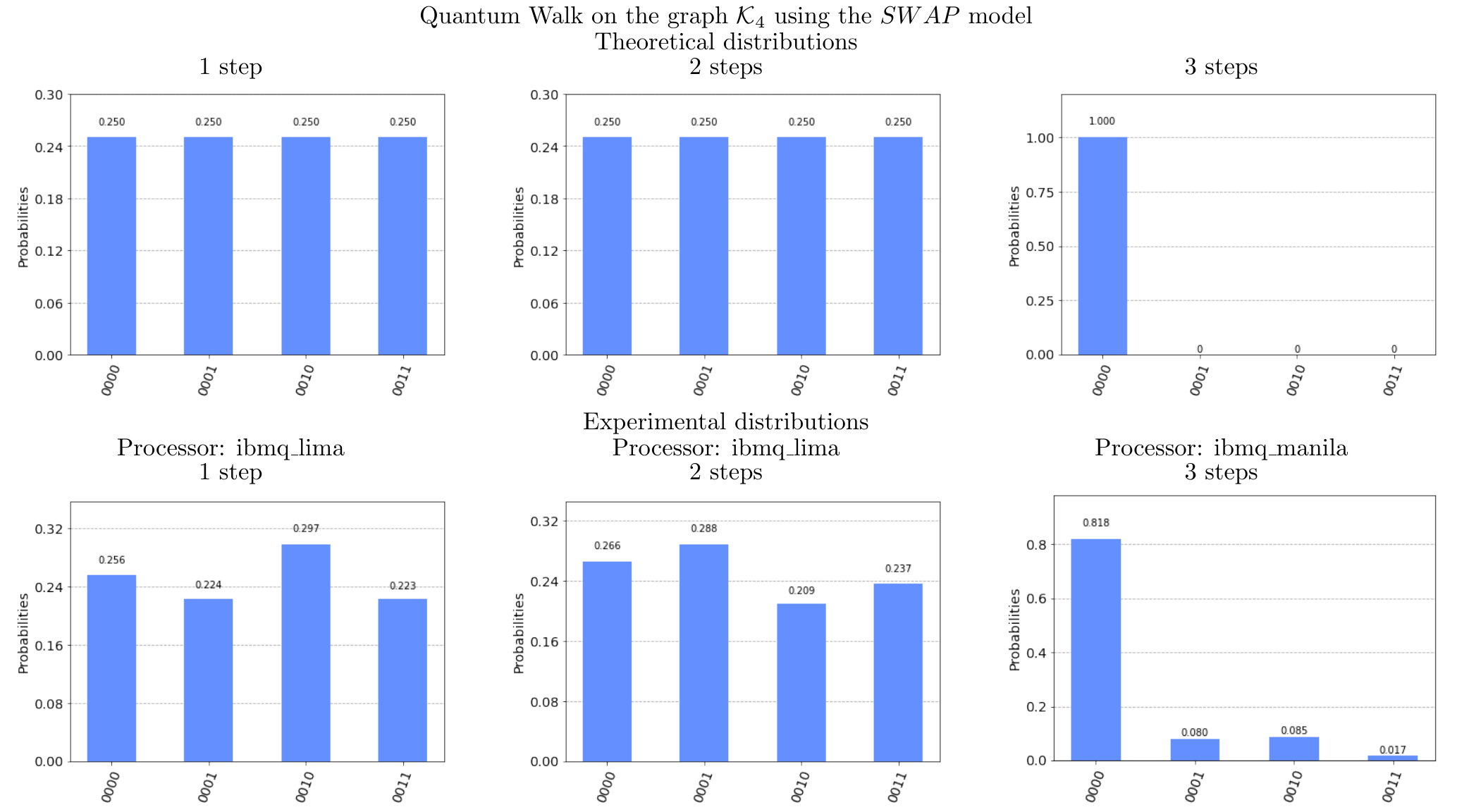}
    }
    \subfigure[]{
    \includegraphics[scale=0.53]{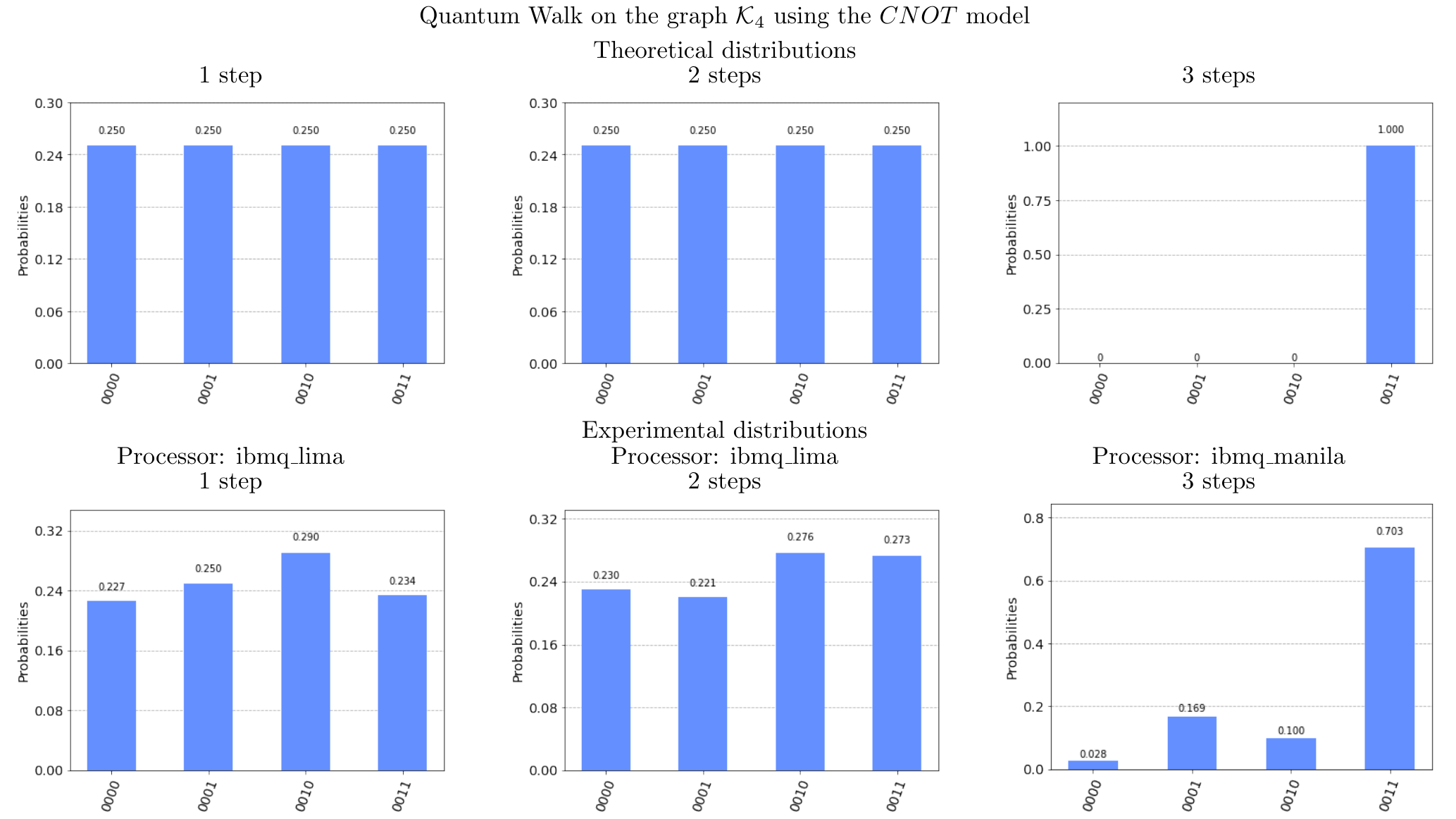}
    }    
    \caption{Comparison between theoretical and best experimental probability distributions of a DTQW on a $\mathcal{K}_4$ graph of 1, 2 and 3 steps. (a) Distribution using the $CNOT$ model and (b) Distribution using the $SWAP$ model. The processor used to obtain each distribution is indicated on top of them. The metrics of each distribution can be found in Tables \ref{K4_Wing_table} and \ref{K4_Wang_table}. These distributions present the best metrics of all the models studied in this work, which is attributed to the simplicity of the circuit in both cases and to the fact that one qubit less was needed to perform the DTQWs in comparison to the other models. Notice that the measured state after a DTQW of three steps is different in both models. Although we perform a DTQW on the same topology, the fact that the $CNOT$ and $SWAP$ models have diagonal and non-diagonal shift operators, respectively, changes the dynamics of the walker}
    \label{k4_best_distros}
\end{figure}

Although the topology is the same for both models, the dynamics of the walker changes depending on the shift operator. The most visible case arises when we compare the third step of both models, while for the $CNOT$ model we obtain the state $|0011\rangle$ after measurement, for the $SWAP$ model we obtain the state $|0000\rangle$ (see Fig. \ref{k4_best_distros}). Additionally, the the walker needs 8 applications of the evolution operator to complete a cycle on the graph using the former model, while with the latter the walker needs only 4 applications. We define the cycle of a DTQW as the number of steps $T$ such that $U^T=I$, and as a consequence the quantum state of a walker $|\psi(t)\rangle$ returns to its initial state. Further simulations using Qiskit, show that the number of steps to complete a cycle in the graph $\mathcal{K}_{2^m}$ remains constant for $m = 1, 2, 4, 8, 16$, which suggests that these might be constant values for any dimension $2^m$.

\subsection{DTQW on the 8-line}

\begin{figure}[b!]
    \centering
    \subfigure[]{
    \includegraphics[scale=0.5]{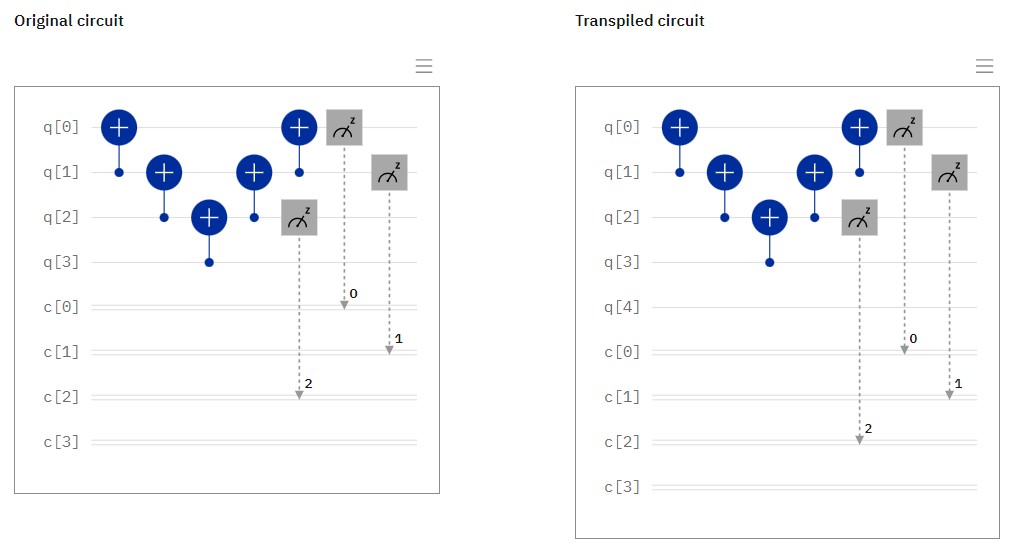}    
    \label{NNA_LRA_original_transpiled_a}    
    }
    \subfigure[]{
    \includegraphics[scale=0.5]{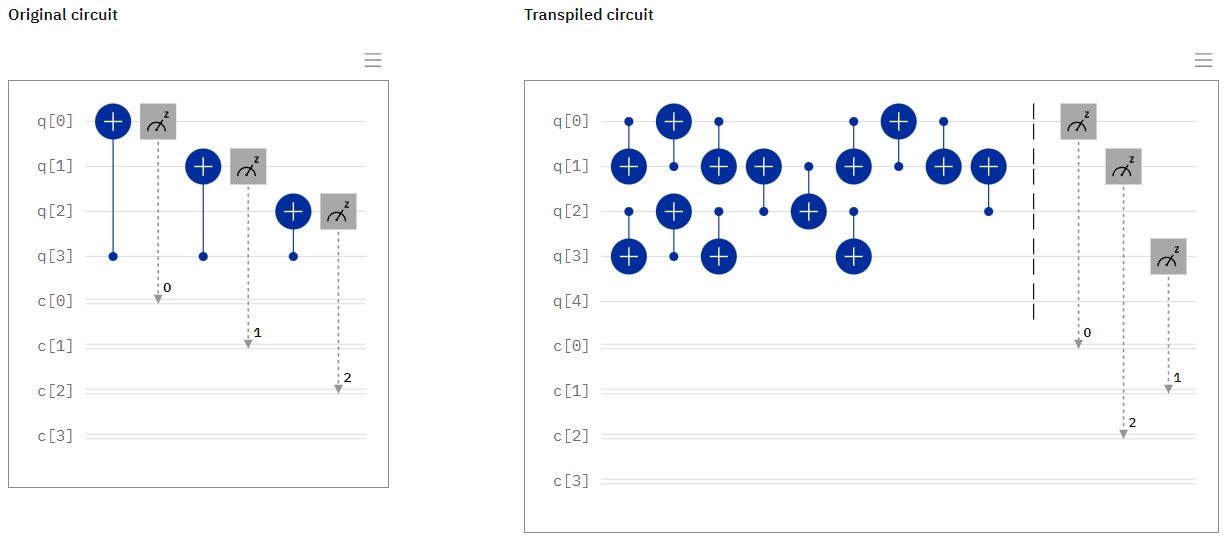}    
    \label{NNA_LRA_original_transpiled_b}    
    }
    \caption{Comparison between the original and transpiled circuit implementations of gate $C_{1,1}(J)$ using (a) NNA and (b) LRA. The quantum circuits were transpiled to the connectivity map of ibmq\_manila, which has a line-like form (see \ref{connectivity_map_a}). Notice from (a) that NNA remains the same after the transpilation process, which indicates it is the optimal arrangement for this type of line-like topologies}
    \label{NNA_LRA_original_transpiled}    
\end{figure}

In the case of the 8-line, two models were considered, for which four qubits were needed. The first model is presented in Fig. \ref{inc-dec_operator_a} and we label it $LRA$ (Long-Range Arrangement); the second model is presented in Fig. \ref{inc-dec_operator_b} and we label it $NNA$ (Nearest-Neighbors Arrangement as proposed in \cite{mottonen}). Comparing these models we want to study if the arrangement of CNOT gates to the sides of the increment operator has an effect on the circuit performance. Specially because NNA is optimal to run in a line-like quantum computer (see Fig. \ref{connectivity_map_a}), as it adjusts perfectly to this topology, while LRA must be transpiled into a completely different sequence of CNOT gates, which increases substantially the total number of gates. This fact can be illustrated from the example presented in Fig. \ref{NNA_LRA_original_transpiled}. Regarding T-like configuration, in principle there is no preferred model.

The $\ell_{1}$ distances between experimental and theoretical distributions are summarized in Tables \ref{nna_8-line_table} and \ref{lra_8-line_table}, where the first table corresponds to the LRA model and the second one to the NNA model. 

\begin{table}[h!]
\caption{$\ell_{1}$ distance between experimental and theoretical distributions for the 8-line using the NNA model. Shorter distances indicate the best experimental performance for each step. ibmq\textunderscore quito has the best overall performance. %From Fig. \ref{NNA_all_distros} we can see that their associated probability distributions resemble the theoretical predictions.  The depth and size of each circuit is available in Table \ref{depth_size_NNA}
} % title of Table
\label{nna_8-line_table} 
\centering % used for centering table
\begin{tabular}{@{}c c c c c@{}} % centered columns (4 columns)
\toprule
Device/Steps & 1 step & 2 steps & 3 steps & Avg\\ % inserts table
%heading
\midrule
ibmq\textunderscore manila & 0.548 & 0.534 & 0.655 & 0.579\\ 
ibmq\textunderscore bogota & 0.457 & 0.597 & \textbf{0.431} & 0.495\\
ibmq\textunderscore quito & \textbf{0.402} & \textbf{0.369} & 0.455 & \textbf{0.409}\\
ibmq\textunderscore lima & 0.548 & 0.534 & 0.655 & 0.579\\ % [1ex] adds vertical space
%\botrule
\end{tabular}
\end{table}
\begin{table}[h!]
\caption{$\ell_{1}$ distance between experimental and theoretical distributions for the 8-line using the LRA model. Shorter distances indicate the best experimental performance for each step. ibmq\textunderscore lima has the best overall performance. %From Fig. \ref{LRA_all_distros} we can see that their associated probability distributions resemble the theoretical predictions. The size and depth of each circuit is available in Table \ref{depth_size_LRA}
} % title of Table
\label{lra_8-line_table} 
\centering % used for centering table
\begin{tabular}{@{}c c c c c@{}} % centered columns (4 columns)
\toprule
Device/Steps & 1 step & 2 steps & 3 steps & Avg \\ % inserts table
%heading
\midrule
ibmq\textunderscore manila & 0.474 & 0.438 & 0.428 & 0.446 \\ 
ibmq\textunderscore bogota & 0.457 & 0.345 & 0.590 & 0.464 \\
ibmq\textunderscore quito & 0.402 & 0.386 & \textbf{0.400} & 0.396\\
ibmq\textunderscore lima & \textbf{0.339} & \textbf{0.337} & 0.456 & \textbf{0.377}\\ % [1ex] adds vertical space
%\botrule
\end{tabular}
\end{table}

\begin{figure}
    \centering
    \includegraphics[scale=0.59]{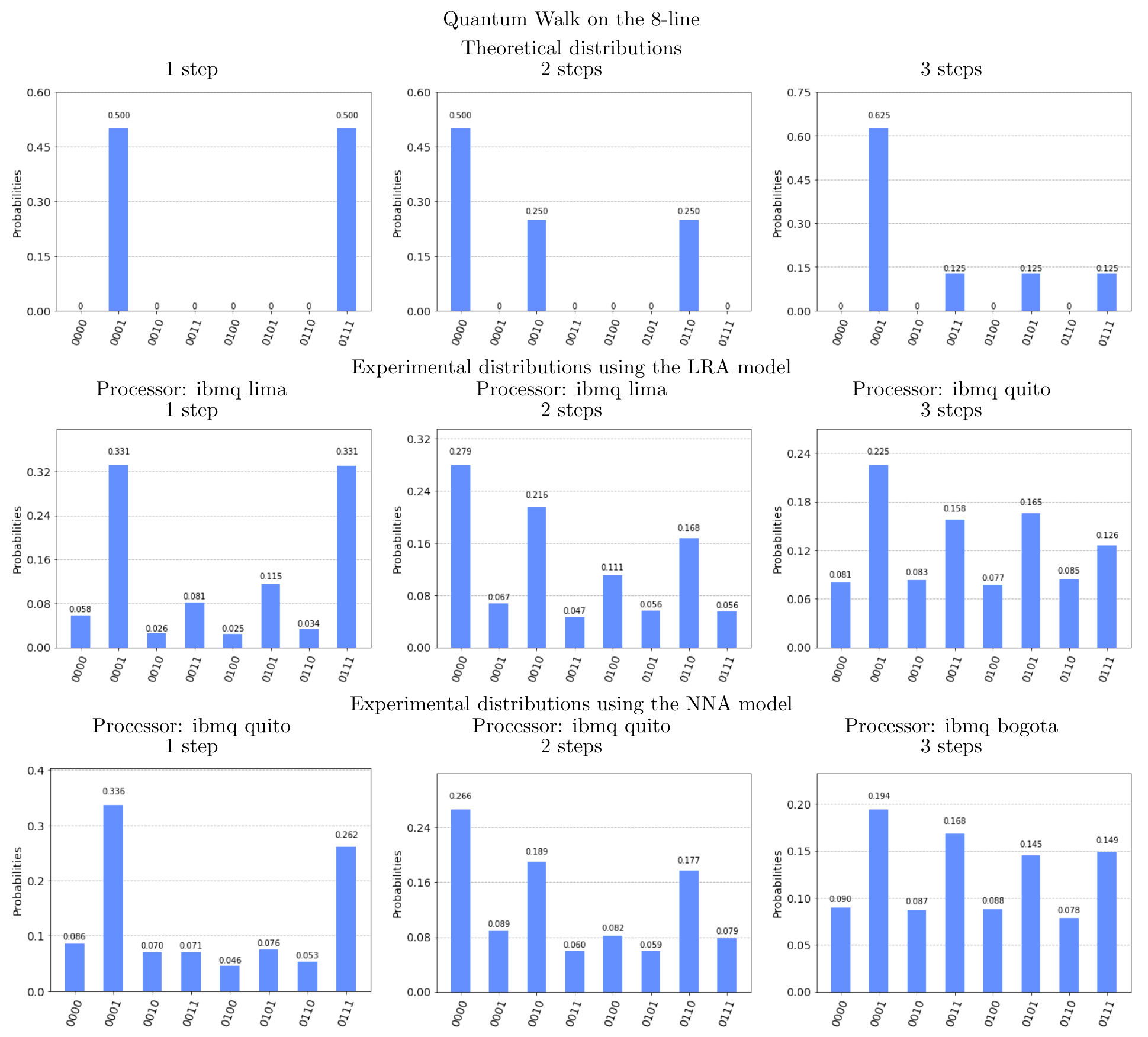}
    \caption{Comparison between theoretical and best experimental probability distributions of a DTQW on an 8-line of 1, 2 and 3 steps. 
 The first row displays the theoretical distributions for each step, while the second and third rows display experimental distributions with the best metrics for each for each step of the LRA and NNA models, respectively. The best metrics are the values in bold in tables \ref{nna_8-line_table} and \ref{lra_8-line_table}. The processor used to obtain each distribution is indicated on top of them. Notice that in all cases, the experimental distributions resemble the theoretical ones, which indicates the circuits were implemented efficiently, hence we have a good performance. This fact is also supported by the low statistical distances for these models reported in the aforementioned tables}
    \label{8-line_best_distros}
\end{figure}

From these tables we can see that the performance of LRA model is better, which we attribute to the depth and size of the original and transpiled circuits. The depth and size of the original one-step circuit for the LRA model are 10 and 10, while the depth and size for its transpiled version with best metric  are 35 and 44, respectively. In the case of the original one-step NNA model, the depth and size are 13 and 14, while the depth and size of its transpiled version with best metric are 43 and 52, respectively. In Fig. \ref{8-line_best_distros} we present the theoretical and experimental distributions for the first three steps on this topology. Notice that regardless of the lower amount of CNOT gates in the NNA model when using a line-like quantum processor (manila and bogota) we still obtain worse results, which suggests that the increment operator has a worse performance in these type of quantum processors.

Attempts to efficiently run the quantum circuit for a DTQW on a line of 16 nodes were done, although the probability distributions in this case were poor given that as we increase the dimensionality of the graph, the number and size of multi-control gates is consequently incremented too, and current NISQ computers have problems maintaining a good fidelity when running circuits with a large number of multi-control gates, as was the case when running 4-qubit instances of Grover's algorithm in \cite{8622457}. This problem was also present when running the circuits for the DTQW on the cycle and hypercube graphs even for low-dimensional instances, thus, in the next sections we decided to replace experimental executions on IBM quantum computers for Qiskit simulations using the Aer simulator in order to prove the functionality of our model.

\subsection{\label{5cycle_section}DTQW on the 5-cycle graph}

\begin{figure}[h!]
\centering
\includegraphics[scale=0.8]{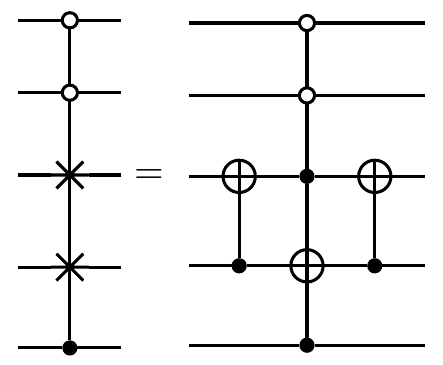}
\caption{Example of the decomposition of a multi-control swap gate in terms of CNOT gates. The left-hand side circuit is a particular case of a multi-cotrol Fredkin gate, which was efficiently decomposed in \cite{fredkin_decomposition, Heese2022representationof}, resulting in the right-hand side circuit}
\label{controlled_swap_decomp}
\end{figure}

In section \ref{cycle_graph_sec} we presented the quantum circuit of the shift operator for a walk on a 5-cycle graph (Fig. \ref{5-cycle_graph_c}), based on CNOT, $C^{n,i}_{m,j}(X)$ and $C^{n,i}_{m,j}(SWAP)$ gates. Qiskit allows us to easily implement the $C^{n,i}_{m,j}(X)$ gate using the \textit{mcx} method. To take advantage of this method, we can use the fact that the $C^{n,i}_{m,j}(SWAP)$ gate, can be decomposed into a set of two CNOT and one $C^{n,i}_{m,j}(X)$ gates, as shown in Fig. \ref{controlled_swap_decomp}.%, then $C^{n,i}_{m,j}(X)$ can be implemented following Fig. \ref{multi-control-z_simplification}. %The decomposed and simplified circuit for a DTQW on a 5-cycle is shown in Fig. \ref{5cycle_decomposed_circuit}.

\begin{figure}[h!]
    \centering
    \includegraphics[scale=0.59]{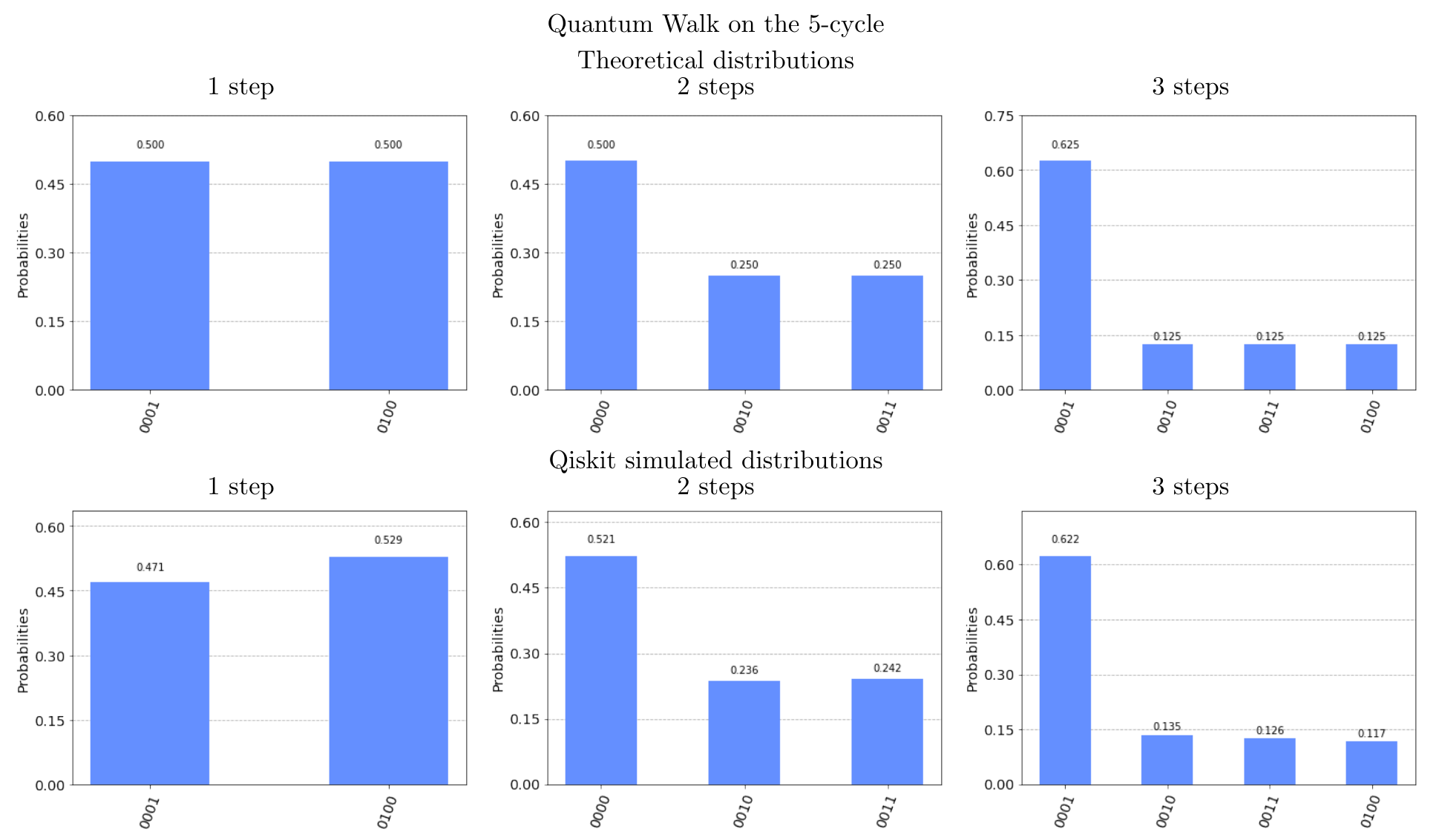}
    \caption{Theoretical and Qiskit-simulated probability distributions of a DTQW of 1, 2 and 3 steps on a 5-cycle graph}
    \label{5cycle_theoretical_distros}
\end{figure}

Fig. \ref{5cycle_theoretical_distros} shows the comparison between Qiskit-simulated and theoretical distributions, for the first, second and third steps of a DTQW on the $5$-cycle, for which we obtained statistical distances of 0.029, 0.0215 and 0.011, respectively.

\subsection{DTQW on the 3-cube with self-loops}

To perform a DTQW on a 3-cube with self-loops, we use the circuit proposed in Fig. \ref{4q-cnot-network_c} for the shift operator, while for the coin operator we use the Grover operator, given that its behavior has been widely studied for this topology in \cite{moore2001quantum, search_alg_portugal_2019, SKB_algoritm, embedded_hypercubes_2014}.

\begin{figure}[h!]
    \centering
    \includegraphics[scale=0.59]{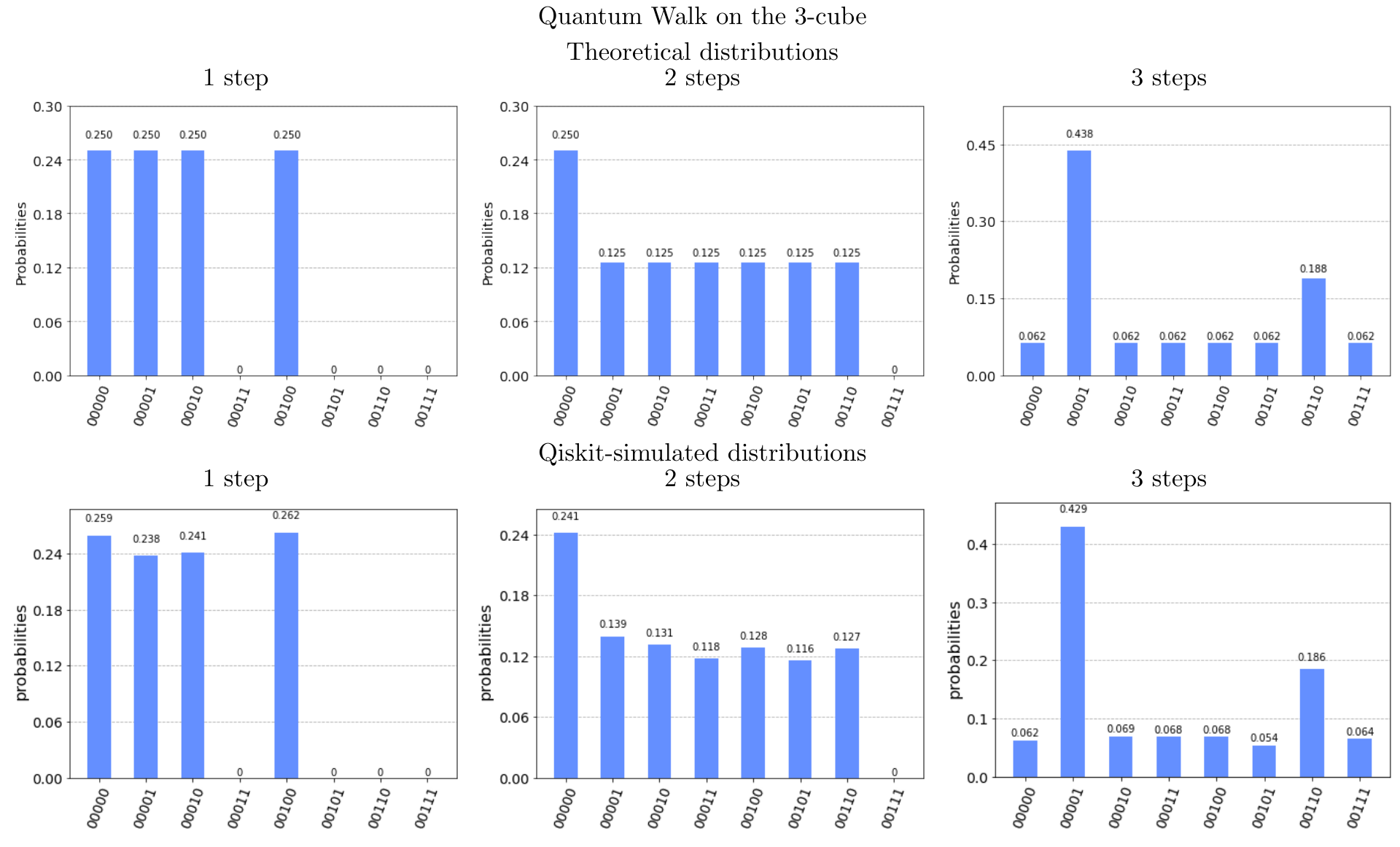}
    \caption{Comparison between theoretical and Qiskit-simulated probability distributions of a DTQW on a 3-cube with self-loops of 1, 2 and 3 steps. The first row displays the theoretical distributions for each step, while the second row displays Qiskit-simulated distributions.}
    \label{3-cube_best_distros}
\end{figure}

Fig. \ref{3-cube_best_distros} shows the comparison between Qiskit-simulated and theoretical distributions, for the first, second and third steps of a DTQW on the $3$-cube, for which we obtained statistical distances of 0.021, 0.025 and 0.0195, respectively.

\section{Conclusion}\label{conclusion_section}

In this paper we have introduced a general method to map evolution operators of DTQWs into quantum circuits, focusing on the case in which the shift operator is a unitary block diagonal matrix, given that this structure allows a direct circuit implementation using fully controlled quantum gates. 

We have used our method to build circuits for graphs commonly presented in the literature of DTQWs, like line, cycle, hypercube and complete graphs. We have shown that these circuits -- some of them previously introduced in the literature -- can be derived from our general method. Furthermore, studying the block diagonal form of the shift operators, we extended number of dimensions in which a DTQW can take place in the case of the hypercube and the cycle graph. For the hypercube, we explained how to modify the circuit proposed by Douglas and Wang \cite{douglas_wang_2009} for a $2^n$-hypercube, to perform a DTQW on a hypercube of any dimension by modifying the number of self-loops in the graph. Additionally for this topology, we provided a way to reduce redundant NOT gates by forcing the pattern created by white and black controls of adjacent multi-control gates to follow two's complement sequence. In the case of the cycle graph, we provided a sequence of gates, which we called adjacent transposition operators, $\mathcal{T}_i$, that allow us the build the shift operator for a cycle graph of any size. Moreover, the adjacent transposition operators might serve as building blocks to construct the quantum circuit associated to any possible permutation matrix, however, due to the fact that all of them are composed of fully controlled gates, they are not suitable to be implemented efficiently in NISQ computers, due to fidelity issues. Finally,  we also used the fact that the decomposition of multi-control quantum gates has been widely studied in order to propose efficient methods to synthesize the circuits, being the most notable case the one for the DTQW on a $2^n$-complete graph, which was reduced from an exponential to a linear number of gates.

In order to prove the reliability of the shunt decomposition method, in section \ref{section5} we ran punctual cases of the quantum circuits studied section \ref{section4}, using both IBM's public quantum computers and Qiskit Aer simulator. The distributions obtained from experimental executions and simulations were compared against analytical results using the $\ell_1$ distance as metric. The topologies we considered were the $\mathcal{K}_4$, 8-line, 5-cycle and 3-cube graphs. However, we were able to obtain short $\ell_1$ distances only for the first two topologies, given that the quantum circuits associated to them were the simplest ones. In the case of the last two topologies, poor results were obtained due to the fact that the circuits for these topologies demand a large number of multi-control not gates, which cannot yet be efficiently run on the quantum computers used. Therefore, we decided to use Qiskit Aer simulations in these cases, obtaining short $\ell_1$ distances in both of them.

Looking to determine whether performance varies from one model to another, in the case of the complete graph, we compared a circuit obtained using the shunt decomposition method ($CNOT$ model) with a circuit whose shift operator is not block diagonal ($SWAP$ model) \cite{douglas_wang_2009}. Both circuits in original form have the same complexity, and even when transpiled, the $SWAP$  transpiled circuit had only a few more quantum gates than the $CNOT$ transpiled circuit (see \cite{github_repo}), which resulted in very short $\ell_1$ distances for both models, as presented in tables \ref{K4_Wing_table} and \ref{K4_Wang_table}. However, we did find out a difference between the models, i.e. in the third step, using the $SWAP$ model, there was $100\%$ probability to find the walker in the state associated to node 0, while for the same step, but using the $CNOT$ model, there was $100\%$ probability for the walker to be in the state associated to node 1 (see Fig. \ref{k4_best_distros}). This provides evidence, that different shift operators for the same topology might lead to different walker dynamics.

Another important aspect to mention is that ibmq\textunderscore quito and ibmq\textunderscore lima were the quantum computers that gave out the largest number of best metrics for specific steps in different models executed experimentally, being ibmq\textunderscore quito the one that had the best average performance in 2 out of 4 models studied. We attribute this to the connectivity of these quantum computers, which is T-like (Fig. \ref{connectivity_map}), rather than other factors like quantum volume which is 16 and 8 for ibmq\textunderscore quito and ibmq\textunderscore lima, respectively, and 32 for both bogota and manila.

Finally, we conclude that the method we developed to map block diagonal unitary operators into quantum circuits can be applied to execute DTQW whose shift operator is constructed based on the shunt decomposition of the adjacency matrix on which the DTQW takes place. Although, the method has experimental limitations in the sense that, in general, it requires a large amount of multi-control gates which induce errors in the execution, we believe it is still be relevant given that is it perfectly suitable to study quantum circuits in an analytical manner and through simulations.

%\backmatter

\section*{Acknowledgments}

Both authors acknowledge the financial support provided by Tecnologico de Monterrey, Escuela de Ingenieria y Ciencias and Consejo Nacional de Ciencia y Tecnología (CONACyT). SEVA acknowledges the support of CONACyT-SNI [SNI number 41594].

%%===========================================================================================%%
%% If you are submitting to one of the Nature Portfolio journals, using the eJP submission   %%
%% system, please include the references within the manuscript file itself. You may do this  %%
%% by copying the reference list from your .bbl file, paste it into the main manuscript .tex %%
%% file, and delete the associated \verb+\bibliography+ commands.                            %%
%%===========================================================================================%%

%\bibliography{sn-bibliography}% common bib file

\begin{thebibliography}{58}
% BibTex style file: bmc-mathphys.bst (version 2.1), 2014-07-24
\ifx \bisbn   \undefined \def \bisbn  #1{ISBN #1}\fi
\ifx \binits  \undefined \def \binits#1{#1}\fi
\ifx \bauthor  \undefined \def \bauthor#1{#1}\fi
\ifx \batitle  \undefined \def \batitle#1{#1}\fi
\ifx \bjtitle  \undefined \def \bjtitle#1{#1}\fi
\ifx \bvolume  \undefined \def \bvolume#1{\textbf{#1}}\fi
\ifx \byear  \undefined \def \byear#1{#1}\fi
\ifx \bissue  \undefined \def \bissue#1{#1}\fi
\ifx \bfpage  \undefined \def \bfpage#1{#1}\fi
\ifx \blpage  \undefined \def \blpage #1{#1}\fi
\ifx \burl  \undefined \def \burl#1{\textsf{#1}}\fi
\ifx \doiurl  \undefined \def \doiurl#1{\url{https://doi.org/#1}}\fi
\ifx \betal  \undefined \def \betal{\textit{et al.}}\fi
\ifx \binstitute  \undefined \def \binstitute#1{#1}\fi
\ifx \binstitutionaled  \undefined \def \binstitutionaled#1{#1}\fi
\ifx \bctitle  \undefined \def \bctitle#1{#1}\fi
\ifx \beditor  \undefined \def \beditor#1{#1}\fi
\ifx \bpublisher  \undefined \def \bpublisher#1{#1}\fi
\ifx \bbtitle  \undefined \def \bbtitle#1{#1}\fi
\ifx \bedition  \undefined \def \bedition#1{#1}\fi
\ifx \bseriesno  \undefined \def \bseriesno#1{#1}\fi
\ifx \blocation  \undefined \def \blocation#1{#1}\fi
\ifx \bsertitle  \undefined \def \bsertitle#1{#1}\fi
\ifx \bsnm \undefined \def \bsnm#1{#1}\fi
\ifx \bsuffix \undefined \def \bsuffix#1{#1}\fi
\ifx \bparticle \undefined \def \bparticle#1{#1}\fi
\ifx \barticle \undefined \def \barticle#1{#1}\fi
%\bibcommenthead
\ifx \bconfdate \undefined \def \bconfdate #1{#1}\fi
\ifx \botherref \undefined \def \botherref #1{#1}\fi
\ifx \url \undefined \def \url#1{\textsf{#1}}\fi
\ifx \bchapter \undefined \def \bchapter#1{#1}\fi
\ifx \bbook \undefined \def \bbook#1{#1}\fi
\ifx \bcomment \undefined \def \bcomment#1{#1}\fi
\ifx \oauthor \undefined \def \oauthor#1{#1}\fi
\ifx \citeauthoryear \undefined \def \citeauthoryear#1{#1}\fi
\ifx \endbibitem  \undefined \def \endbibitem {}\fi
\ifx \bconflocation  \undefined \def \bconflocation#1{#1}\fi
\ifx \arxivurl  \undefined \def \arxivurl#1{\textsf{#1}}\fi
\csname PreBibitemsHook\endcsname

%%% 1
\bibitem{lawler_limic_2010}
\begin{botherref}
\oauthor{\bsnm{Lawler}, \binits{G.F.}},
\oauthor{\bsnm{Limic}, \binits{V.}}:
Random walk: A modern introduction
(2010).
\doiurl{10.1017/CBO9780511750854}
\end{botherref}
\endbibitem

%%% 2
\bibitem{10.2307/27851819}
\begin{barticle}
\bauthor{\bsnm{Weiss}, \binits{G.H.}}:
\batitle{Random walks and their applications: Widely used as mathematical
  models, random walks play an important role in several areas of physics,
  chemistry, and biology}.
\bjtitle{American Scientist}
\bvolume{71}(\bissue{1}),
\bfpage{65}--\blpage{71}
(\byear{1983})
\end{barticle}
\endbibitem

%%% 3
\bibitem{codling_plank_benhamou_2008}
\begin{barticle}
\bauthor{\bsnm{Codling}, \binits{E.A.}},
\bauthor{\bsnm{Plank}, \binits{M.J.}},
\bauthor{\bsnm{Benhamou}, \binits{S.}}:
\batitle{Random walk models in biology}.
\bjtitle{Journal of The Royal Society Interface}
\bvolume{5}(\bissue{25}),
\bfpage{813}--\blpage{834}
(\byear{2008}).
\doiurl{10.1098/rsif.2008.0014}
\end{barticle}
\endbibitem

%%% 4
\bibitem{7299008}
\begin{bchapter}
\bauthor{\bsnm{Lee}, \binits{C.}},
\bauthor{\bsnm{Jang}, \binits{W.-D.}},
\bauthor{\bsnm{Sim}, \binits{J.-Y.}},
\bauthor{\bsnm{Kim}, \binits{C.-S.}}:
\bctitle{Multiple random walkers and their application to image
  cosegmentation}.
In: \bbtitle{2015 IEEE Conference on Computer Vision and Pattern Recognition
  (CVPR)},
pp. \bfpage{3837}--\blpage{3845}
(\byear{2015}).
\doiurl{10.1109/CVPR.2015.7299008}
\end{bchapter}
\endbibitem

%%% 5
\bibitem{abakah_alagidede_mensah_ohene-asare_2018}
\begin{barticle}
\bauthor{\bsnm{Abakah}, \binits{E.J.}},
\bauthor{\bsnm{Alagidede}, \binits{P.}},
\bauthor{\bsnm{Mensah}, \binits{L.}},
\bauthor{\bsnm{Ohene-Asare}, \binits{K.}}:
\batitle{Non-linear approach to random walk test in selected african
  countries}.
\bjtitle{International Journal of Managerial Finance}
\bvolume{14}(\bissue{3}),
\bfpage{362}--\blpage{376}
(\byear{2018}).
\doiurl{10.1108/ijmf-10-2017-0235}
\end{barticle}
\endbibitem

%%% 6
\bibitem{doi:10.1063/1.4903129}
\begin{barticle}
\bauthor{\bsnm{Kendon}, \binits{V.}}:
\batitle{Quantum walk computation}.
\bjtitle{AIP Conference Proceedings}
\bvolume{1633}(\bissue{1}),
\bfpage{177}--\blpage{179}
(\byear{2014})
{\href{https://arxiv.org/abs/https://aip.scitation.org/doi/pdf/10.1063/1.4903129}{{https://aip.scitation.org/doi/pdf/10.1063/1.4903129}}}.
\doiurl{10.1063/1.4903129}
\end{barticle}
\endbibitem

%%% 7
\bibitem{krovi_magniez_ozols_roland_2015}
\begin{barticle}
\bauthor{\bsnm{Krovi}, \binits{H.}},
\bauthor{\bsnm{Magniez}, \binits{F.}},
\bauthor{\bsnm{Ozols}, \binits{M.}},
\bauthor{\bsnm{Roland}, \binits{J.}}:
\batitle{Quantum walks can find a marked element on any graph}.
\bjtitle{Algorithmica}
\bvolume{74}(\bissue{2}),
\bfpage{851}--\blpage{907}
(\byear{2015}).
\doiurl{10.1007/s00453-015-9979-8}
\end{barticle}
\endbibitem

%%% 8
\bibitem{aharonov2002quantum}
\begin{botherref}
\oauthor{\bsnm{Aharonov}, \binits{D.}},
\oauthor{\bsnm{Ambainis}, \binits{A.}},
\oauthor{\bsnm{Kempe}, \binits{J.}},
\oauthor{\bsnm{Vazirani}, \binits{U.}}:
Quantum Walks On Graphs
(2002)
\end{botherref}
\endbibitem

%%% 9
\bibitem{quantum_decision_trees1998}
\begin{barticle}
\bauthor{\bsnm{Farhi}, \binits{E.}},
\bauthor{\bsnm{Gutmann}, \binits{S.}}:
\batitle{Quantum computation and decision trees}.
\bjtitle{Physical Review A}
\bvolume{58}(\bissue{2}),
\bfpage{915}--\blpage{928}
(\byear{1998}).
\doiurl{10.1103/physreva.58.915}
\end{barticle}
\endbibitem

%%% 10
\bibitem{chandrashekar2010discretetime}
\begin{botherref}
\oauthor{\bsnm{Chandrashekar}, \binits{C.M.}}:
Discrete-Time Quantum Walk - Dynamics and Applications
(2010)
\end{botherref}
\endbibitem

%%% 11
\bibitem{szegedy_2004}
\begin{barticle}
\bauthor{\bsnm{Szegedy}, \binits{M.}}:
\batitle{Quantum speed-up of markov chain based algorithms}.
\bjtitle{45th Annual IEEE Symposium on Foundations of Computer Science}
(\byear{2004}).
\doiurl{10.1109/focs.2004.53}
\end{barticle}
\endbibitem

%%% 12
\bibitem{PhysRevA.93.062335}
\begin{barticle}
\bauthor{\bsnm{Portugal}, \binits{R.}}:
\batitle{Staggered quantum walks on graphs}.
\bjtitle{Phys. Rev. A}
\bvolume{93},
\bfpage{062335}
(\byear{2016}).
\doiurl{10.1103/PhysRevA.93.062335}
\end{barticle}
\endbibitem

%%% 13
\bibitem{seva2012}
\begin{barticle}
\bauthor{\bsnm{Venegas-Andraca}, \binits{S..}}:
\batitle{Quantum walk: a comprehensive review}.
\bjtitle{Quantum Information Processing}
\bvolume{11}(\bissue{5}),
\bfpage{1015}--\blpage{1106}
(\byear{2012})
\end{barticle}
\endbibitem

%%% 14
\bibitem{yang_pan_sun_xu_2015}
\begin{botherref}
\oauthor{\bsnm{Yang}, \binits{Y.-G.}},
\oauthor{\bsnm{Pan}, \binits{Q.-X.}},
\oauthor{\bsnm{Sun}, \binits{S.-J.}},
\oauthor{\bsnm{Xu}, \binits{P.}}:
Novel image encryption based on quantum walks.
Scientific Reports
\textbf{5}(1)
(2015).
\doiurl{10.1038/srep07784}
\end{botherref}
\endbibitem

%%% 15
\bibitem{public_key_encryption_2015}
\begin{barticle}
\bauthor{\bsnm{Vlachou}, \binits{C.}},
\bauthor{\bsnm{Rodrigues}, \binits{J.}},
\bauthor{\bsnm{Mateus}, \binits{P.}},
\bauthor{\bsnm{Paunković}, \binits{N.}},
\bauthor{\bsnm{Souto}, \binits{A.}}:
\batitle{Quantum walk public-key cryptographic system}.
\bjtitle{International Journal of Quantum Information}
\bvolume{13}(\bissue{07}),
\bfpage{1550050}
(\byear{2015}).
\doiurl{10.1142/s0219749915500501}
\end{barticle}
\endbibitem

%%% 16
\bibitem{SKB_algoritm}
\begin{botherref}
\oauthor{\bsnm{Shenvi}, \binits{N.}},
\oauthor{\bsnm{Kempe}, \binits{J.}},
\oauthor{\bsnm{Whaley}, \binits{K.B.}}:
Quantum random-walk search algorithm.
Physical Review A
\textbf{67}(5)
(2003).
\doiurl{10.1103/physreva.67.052307}
\end{botherref}
\endbibitem

%%% 17
\bibitem{bezerra_lugao_portugal_2021}
\begin{botherref}
\oauthor{\bsnm{Bezerra}, \binits{G.A.}},
\oauthor{\bsnm{Lugão}, \binits{P.H.}},
\oauthor{\bsnm{Portugal}, \binits{R.}}:
Quantum-walk-based search algorithms with multiple marked vertices.
Physical Review A
\textbf{103}(6)
(2021).
\doiurl{10.1103/physreva.103.062202}
\end{botherref}
\endbibitem

%%% 18
\bibitem{dernbach_mohseni-kabir_pal_gepner_towsley_2019}
\begin{botherref}
\oauthor{\bsnm{Dernbach}, \binits{S.}},
\oauthor{\bsnm{Mohseni-Kabir}, \binits{A.}},
\oauthor{\bsnm{Pal}, \binits{S.}},
\oauthor{\bsnm{Gepner}, \binits{M.}},
\oauthor{\bsnm{Towsley}, \binits{D.}}:
Quantum walk neural networks with feature dependent coins.
Applied Network Science
\textbf{4}(1)
(2019).
\doiurl{10.1007/s41109-019-0188-2}
\end{botherref}
\endbibitem

%%% 19
\bibitem{8923910}
\begin{bchapter}
\bauthor{\bparticle{de} \bsnm{Souza}, \binits{L.S.}},
\bauthor{\bparticle{de} \bsnm{Carvalho}, \binits{J.H.A.}},
\bauthor{\bsnm{Ferreira}, \binits{T.A.E.}}:
\bctitle{Quantum walk to train a classical artificial neural network}.
In: \bbtitle{2019 8th Brazilian Conference on Intelligent Systems (BRACIS)},
pp. \bfpage{836}--\blpage{841}
(\byear{2019}).
\doiurl{10.1109/BRACIS.2019.00149}
\end{bchapter}
\endbibitem

%%% 20
\bibitem{paparo_martin-delgado_2012}
\begin{botherref}
\oauthor{\bsnm{Paparo}, \binits{G.D.}},
\oauthor{\bsnm{Martin-Delgado}, \binits{M.A.}}:
Google in a quantum network.
Scientific Reports
\textbf{2}(1)
(2012).
\doiurl{10.1038/srep00444}
\end{botherref}
\endbibitem

%%% 21
\bibitem{pagerank_chawla}
\begin{botherref}
\oauthor{\bsnm{Chawla}, \binits{P.}},
\oauthor{\bsnm{Mangal}, \binits{R.}},
\oauthor{\bsnm{Chandrashekar}, \binits{C.M.}}:
Discrete-time quantum walk algorithm for ranking nodes on a network.
Quantum Information Processing
\textbf{19}(5)
(2020).
\doiurl{10.1007/s11128-020-02650-4}
\end{botherref}
\endbibitem

%%% 22
\bibitem{PhysRevA.78.012310}
\begin{barticle}
\bauthor{\bsnm{Tulsi}, \binits{A.}}:
\batitle{Faster quantum-walk algorithm for the two-dimensional spatial search}.
\bjtitle{Phys. Rev. A}
\bvolume{78},
\bfpage{012310}
(\byear{2008}).
\doiurl{10.1103/PhysRevA.78.012310}
\end{barticle}
\endbibitem

%%% 23
\bibitem{preskill_2018}
\begin{barticle}
\bauthor{\bsnm{Preskill}, \binits{J.}}:
\batitle{Quantum computing in the nisq era and beyond}.
\bjtitle{Quantum}
\bvolume{2},
\bfpage{79}
(\byear{2018}).
\doiurl{10.22331/q-2018-08-06-79}
\end{barticle}
\endbibitem

%%% 24
\bibitem{PhysRevLett.104.050502}
\begin{barticle}
\bauthor{\bsnm{Schreiber}, \binits{A.}},
\bauthor{\bsnm{Cassemiro}, \binits{K.N.}},
\bauthor{\bparticle{Poto\ifmmode~\check{c}\else} \bsnm{\v{c}\fi{}ek},
  \binits{V.}},
\bauthor{\bsnm{G\'abris}, \binits{A.}},
\bauthor{\bsnm{Mosley}, \binits{P.J.}},
\bauthor{\bsnm{Andersson}, \binits{E.}},
\bauthor{\bsnm{Jex}, \binits{I.}},
\bauthor{\bsnm{Silberhorn}, \binits{C.}}:
\batitle{Photons walking the line: A quantum walk with adjustable coin
  operations}.
\bjtitle{Phys. Rev. Lett.}
\bvolume{104},
\bfpage{050502}
(\byear{2010}).
\doiurl{10.1103/PhysRevLett.104.050502}
\end{barticle}
\endbibitem

%%% 25
\bibitem{PhysRevLett.104.153602}
\begin{barticle}
\bauthor{\bsnm{Broome}, \binits{M.A.}},
\bauthor{\bsnm{Fedrizzi}, \binits{A.}},
\bauthor{\bsnm{Lanyon}, \binits{B.P.}},
\bauthor{\bsnm{Kassal}, \binits{I.}},
\bauthor{\bsnm{Aspuru-Guzik}, \binits{A.}},
\bauthor{\bsnm{White}, \binits{A.G.}}:
\batitle{Discrete single-photon quantum walks with tunable decoherence}.
\bjtitle{Phys. Rev. Lett.}
\bvolume{104},
\bfpage{153602}
(\byear{2010}).
\doiurl{10.1103/PhysRevLett.104.153602}
\end{barticle}
\endbibitem

%%% 26
\bibitem{shakeel_2020}
\begin{botherref}
\oauthor{\bsnm{Shakeel}, \binits{A.}}:
Efficient and scalable quantum walk algorithms via the quantum fourier
  transform.
Quantum Information Processing
\textbf{19}(9)
(2020).
\doiurl{10.1007/s11128-020-02834-y}
\end{botherref}
\endbibitem

%%% 27
\bibitem{PhysRevA.103.022408}
\begin{barticle}
\bauthor{\bsnm{Georgopoulos}, \binits{K.}},
\bauthor{\bsnm{Emary}, \binits{C.}},
\bauthor{\bsnm{Zuliani}, \binits{P.}}:
\batitle{Comparison of quantum-walk implementations on noisy intermediate-scale
  quantum computers}.
\bjtitle{Phys. Rev. A}
\bvolume{103},
\bfpage{022408}
(\byear{2021}).
\doiurl{10.1103/PhysRevA.103.022408}
\end{barticle}
\endbibitem

%%% 28
\bibitem{acasiete_agostini_moqadam_portugal_2020}
\begin{botherref}
\oauthor{\bsnm{Acasiete}, \binits{F.}},
\oauthor{\bsnm{Agostini}, \binits{F.P.}},
\oauthor{\bsnm{Moqadam}, \binits{J.K.}},
\oauthor{\bsnm{Portugal}, \binits{R.}}:
Implementation of quantum walks on ibm quantum computers.
Quantum Information Processing
\textbf{19}(12)
(2020).
\doiurl{10.1007/s11128-020-02938-5}
\end{botherref}
\endbibitem

%%% 29
\bibitem{Balu_2018}
\begin{barticle}
\bauthor{\bsnm{Balu}, \binits{R.}},
\bauthor{\bsnm{Castillo}, \binits{D.}},
\bauthor{\bsnm{Siopsis}, \binits{G.}}:
\batitle{Physical realization of topological quantum walks on {IBM}-q and
  beyond}.
\bjtitle{Quantum Science and Technology}
\bvolume{3}(\bissue{3}),
\bfpage{035001}
(\byear{2018}).
\doiurl{10.1088/2058-9565/aab823}
\end{barticle}
\endbibitem

%%% 30
\bibitem{tang_lin_feng_chen_gao_sun_wang_lai_xu_wang_2018}
\begin{botherref}
\oauthor{\bsnm{Tang}, \binits{H.}},
\oauthor{\bsnm{Lin}, \binits{X.-F.}},
\oauthor{\bsnm{Feng}, \binits{Z.}},
\oauthor{\bsnm{Chen}, \binits{J.-Y.}},
\oauthor{\bsnm{Gao}, \binits{J.}},
\oauthor{\bsnm{Sun}, \binits{K.}},
\oauthor{\bsnm{Wang}, \binits{C.-Y.}},
\oauthor{\bsnm{Lai}, \binits{P.-C.}},
\oauthor{\bsnm{Xu}, \binits{X.-Y.}},
\oauthor{\bsnm{Wang}, \binits{Y.}},
\oauthor{\bparticle{et} \bsnm{al.}}:
Experimental two-dimensional quantum walk on a photonic chip.
Science Advances
\textbf{4}(5)
(2018).
\doiurl{10.1126/sciadv.aat3174}
\end{botherref}
\endbibitem

%%% 31
\bibitem{qiang_loke_montanaro_aungskunsiri_zhou_obrien_wang_matthews_2016}
\begin{botherref}
\oauthor{\bsnm{Qiang}, \binits{X.}},
\oauthor{\bsnm{Loke}, \binits{T.}},
\oauthor{\bsnm{Montanaro}, \binits{A.}},
\oauthor{\bsnm{Aungskunsiri}, \binits{K.}},
\oauthor{\bsnm{Zhou}, \binits{X.}},
\oauthor{\bsnm{O’Brien}, \binits{J.L.}},
\oauthor{\bsnm{Wang}, \binits{J.B.}},
\oauthor{\bsnm{Matthews}, \binits{J.C.}}:
Efficient quantum walk on a quantum processor.
Nature Communications
\textbf{7}(1)
(2016).
\doiurl{10.1038/ncomms11511}
\end{botherref}
\endbibitem

%%% 32
\bibitem{Jiao:21}
\begin{barticle}
\bauthor{\bsnm{Jiao}, \binits{Z.-Q.}},
\bauthor{\bsnm{Gao}, \binits{J.}},
\bauthor{\bsnm{Zhou}, \binits{W.-H.}},
\bauthor{\bsnm{Wang}, \binits{X.-W.}},
\bauthor{\bsnm{Ren}, \binits{R.-J.}},
\bauthor{\bsnm{Xu}, \binits{X.-Y.}},
\bauthor{\bsnm{Qiao}, \binits{L.-F.}},
\bauthor{\bsnm{Wang}, \binits{Y.}},
\bauthor{\bsnm{Jin}, \binits{X.-M.}}:
\batitle{Two-dimensional quantum walks of correlated photons}.
\bjtitle{Optica}
\bvolume{8}(\bissue{9}),
\bfpage{1129}--\blpage{1135}
(\byear{2021}).
\doiurl{10.1364/OPTICA.425879}
\end{barticle}
\endbibitem

%%% 33
\bibitem{GODSIL2019181}
\begin{barticle}
\bauthor{\bsnm{Godsil}, \binits{C.}},
\bauthor{\bsnm{Zhan}, \binits{H.}}:
\batitle{Discrete-time quantum walks and graph structures}.
\bjtitle{Journal of Combinatorial Theory, Series A}
\bvolume{167},
\bfpage{181}--\blpage{212}
(\byear{2019}).
\doiurl{10.1016/j.jcta.2019.05.003}
\end{barticle}
\endbibitem

%%% 34
\bibitem{montanaro_2007}
\begin{barticle}
\bauthor{\bsnm{Montanaro}, \binits{A.}}:
\batitle{Quantum walks on directed graphs}.
\bjtitle{Quantum Information and Computation}
\bvolume{7}(\bissue{1\&2}),
\bfpage{93}--\blpage{102}
(\byear{2007}).
\doiurl{10.26421/qic7.1-2-5}
\end{barticle}
\endbibitem

%%% 35
\bibitem{carnia_adjacency_clycle_complete}
\begin{barticle}
\bauthor{\bsnm{Carnia}, \binits{E.}},
\bauthor{\bsnm{Suyudi}, \binits{M.}},
\bauthor{\bsnm{Aisah}, \binits{I.}},
\bauthor{\bsnm{Supriatna}, \binits{A.K.}}:
\batitle{A review on eigen values of adjacency matrix of graph with cliques}.
\bjtitle{AIP Conference Proceedings}
(\byear{2017}).
\doiurl{10.1063/1.4995116}
\end{barticle}
\endbibitem

%%% 36
\bibitem{li_yang_increment}
\begin{barticle}
\bauthor{\bsnm{Li}, \binits{X.}},
\bauthor{\bsnm{Yang}, \binits{G.}},
\bauthor{\bsnm{Torres}, \binits{C.L.}},
\bauthor{\bsnm{Zheng}, \binits{D.}},
\bauthor{\bsnm{Wang}, \binits{K.L.}}:
\batitle{A class of efficient quantum incrementer gates for quantum circuit
  synthesis}.
\bjtitle{International Journal of Modern Physics B}
\bvolume{28}(\bissue{01}),
\bfpage{1350191}
(\byear{2013}).
\doiurl{10.1142/s0217979213501919}
\end{barticle}
\endbibitem

%%% 37
\bibitem{golub_2013}
\begin{botherref}
\oauthor{\bsnm{Golub}, \binits{G.}}
\oauthor{\bsnm{Van Loan}, \binits{C.}}:
Matrix computations, 4th Edition. Johns Hopkins Studies in the Mathematical Sciences, Johns Hopkins University Press, Baltimore (2013).
\end{botherref}
\endbibitem

%%% 38
\bibitem{mottonen}
\begin{botherref}
\oauthor{\bsnm{Mottonen}, \binits{M.}},
\oauthor{\bsnm{Vartiainen}, \binits{J.J.}}:
\batitle{Decompositions of general quantum gates}. 
\bjtitle{arXiv}
(\byear{2005}).
\doiurl{10.48550/ARXIV.QUANT-PH/0504100}.
{\href{https://arxiv.org/abs/quant-ph/0504100}{{https://arxiv.org/abs/quant-ph/0504100}}}
\end{botherref}
\endbibitem

%%% 39
\bibitem{qc_paulesina}
\begin{botherref}
\oauthor{\bsnm{Tucci}, \binits{R.R.}}:
\batitle{QC Paulinesia}. 
\bjtitle{arXiv}
(\byear{2004}).
\doiurl{https://doi.org/10.1007/978-1-4}.
{\href{https://arxiv.org/abs/quant-ph/0407215}{{https://arxiv.org/abs/quant-ph/0407215}}}
\end{botherref}
\endbibitem

%%% 40
\bibitem{unitary_decomposition_li}
\begin{barticle}
\bauthor{\bsnm{LI}, \binits{C.-K.}},
\bauthor{\bsnm{Roberts}, \binits{R.}},
\bauthor{\bsnm{Yin}, \binits{X.}}:
\batitle{Decomposition of unitary matrices and quantum gates}.
\bjtitle{International Journal of Quantum Information}
\bvolume{11}(\bissue{01}),
\bfpage{1350015}
(\byear{2013}).
\doiurl{10.1142/s0219749913500159}
\end{barticle}
\endbibitem

%%% 41
\bibitem{rotman_1995}
\begin{botherref}
\oauthor{\bsnm{Rotman}, \binits{J.J.}}:
An introduction to the theory of groups, 4th Edition. Springer New York, NY, Baltimore (1994).
\doiurl{https://doi.org/10.1007/978-1-4612-4176-8}.
\end{botherref}
\endbibitem

%%% 42
\bibitem{DLMF}
\begin{botherref}
\oauthor{\bsnm{Olver}, \binits{F.W.J.}},
\oauthor{\bparticle{et} \bsnm{al.}}:
\batitle{NIST Digital Library of Mathematical Functions}. 
\bjtitle{http://dlmf.nist.gov/},
Release 1.0.26 of 2020-03-15.
Section 26.13 Permutations: Cycle Notation
(\byear{2020}).
{\href{http://dlmf.nist.gov/}{{http://dlmf.nist.gov/}}}
\end{botherref}
\endbibitem

%%% 43
\bibitem{florkowski_hypcube_ad_mat}
\begin{botherref}
\oauthor{\bsnm{Florkowski}, \binits{S.F.}}:
Spectral graph theory of the hypercube.
PhD thesis,
Naval Postgraduate School
(2008)
\end{botherref}
\endbibitem

%%% 44
\bibitem{douglas_wang_2009}
\begin{botherref}
\oauthor{\bsnm{Douglas}, \binits{B.L.}},
\oauthor{\bsnm{Wang}, \binits{J.B.}}:
Efficient quantum circuit implementation of quantum walks.
Physical Review A
\textbf{79}(5)
(2009).
\doiurl{10.1103/physreva.79.052335}
\end{botherref}
\endbibitem

%%% 45
\bibitem{daraeizadeh_kumar_2014}
\begin{botherref}
\oauthor{\bsnm{Daraeizadeh}, \binits{S.}},
\oauthor{\bsnm{Kumar}, \binits{P.}}:
Efficient implementation of multi-control toffoli gates in linear nearest
  neighbor arrays.
PhD thesis,
Wichita State University
(2014)
\end{botherref}
\endbibitem

%%% 46
\bibitem{rahman_rice_2014}
\begin{botherref}
\oauthor{\bsnm{Rahman}, \binits{M.Z.}},
\oauthor{\bsnm{Rice}, \binits{J.E.}}:
Templates for positive and negative control toffoli networks.
Reversible Computation,
125--136
(2014).
\doiurl{10.1007/978-3-319-08494-7_10}
\end{botherref}
\endbibitem

%%% 47
\bibitem{cheng_guan_wang_zhu_2012}
\begin{barticle}
\bauthor{\bsnm{Cheng}, \binits{X.}},
\bauthor{\bsnm{Guan}, \binits{Z.}},
\bauthor{\bsnm{Wang}, \binits{W.}},
\bauthor{\bsnm{Zhu}, \binits{L.}}:
\batitle{A simplification algorithm for reversible logic network of
  positive/negative control gates}.
\bjtitle{2012 9th International Conference on Fuzzy Systems and Knowledge
  Discovery}
(\byear{2012}).
\doiurl{10.1109/fskd.2012.6233837}
\end{barticle}
\endbibitem

%%% 48
\bibitem{arabzadeh_saeedi_zamani_2010}
\begin{barticle}
\bauthor{\bsnm{Arabzadeh}, \binits{M.}},
\bauthor{\bsnm{Saeedi}, \binits{M.}},
\bauthor{\bsnm{Zamani}, \binits{M.S.}}:
\batitle{Rule-based optimization of reversible circuits}.
\bjtitle{2010 15th Asia and South Pacific Design Automation Conference
  (ASP-DAC)}
(\byear{2010}).
\doiurl{10.1109/aspdac.2010.5419685}
\end{barticle}
\endbibitem

%%% 49
\bibitem{ibm_quantum}
\begin{botherref}
IBM quantum.
\url{https://quantum-computing.ibm.com/}
\end{botherref}
\endbibitem

%%% 50
\bibitem{ibm_quantum_processors}
\begin{botherref}
IBM quantum processor types.
\url{https://quantum-computing.ibm.com/composer/docs/iqx/manage/systems/processors}
\end{botherref}
\endbibitem

%%% 51
\bibitem{github_repo}
\begin{botherref}
\oauthor{\bsnm{Wing}, \binits{A.}}:
Allanwing-QC/quantum-walks-via-shunt-decomposition-circuits
(2022).
\url{https://github.com/allanwing-qc/Quantum-Walks-via-Shunt-Decomposition-Circuits}
\end{botherref}
\endbibitem


%%% 52
\bibitem{8622457}
\begin{barticle}
\oauthor{\bsnm{Mandviwalla}, \binits{A.}},
\oauthor{\bsnm{Ohshiro}, \binits{K.}},
\oauthor{\bsnm{Ji}, \binits{B.}}:
\batitle{Implementing Grover’s Algorithm on the IBM Quantum Computers}. 
\bjtitle{2018 IEEE International Conference on Big Data (Big Data)},
2531-2537
(\byear{2018}).
\doiurl{10.1109/BigData.2018.8622457}
\end{barticle}
\endbibitem

%%% 53
\bibitem{fredkin_decomposition}
\begin{botherref}
\oauthor{\bsnm{Liu}, \binits{J.}},
\oauthor{\bsnm{Bello}, \binits{L.}},
\oauthor{\bsnm{Zhou}, \binits{H.}}:
\batitle{Relaxed Peephole Optimization: A Novel Compiler Optimization for Quantum Circuits}. 
\bjtitle{arXiv}
(\byear{2020}).
\doiurl{10.48550/ARXIV.2012.07711}.
{\href{https://arxiv.org/abs/2012.07711}{{https://arxiv.org/abs/2012.07711}}}
\end{botherref}
\endbibitem

%%% 54
\bibitem{Heese2022representationof}
\begin{barticle}
\oauthor{\bsnm{Heese}, \binits{R.}},
\oauthor{\bsnm{Bickert}, \binits{P.}},
\oauthor{\bsnm{Niederle}, \binits{A.E.}}:
\batitle{Representation of binary classification trees with binary features by quantum circuits}. 
\bjtitle{Quantum} 
\bvolume{6}, 676
(\byear{2022}).
\doiurl{10.22331/q-2022-03-30-676}
\end{barticle}
\endbibitem

%%% 55
\bibitem{moore2001quantum}
\begin{botherref}
\oauthor{\bsnm{Moore}, \binits{C.}},
\oauthor{\bsnm{Russell}, \binits{A.}}:
Quantum Walks on the Hypercube
(2001)
\end{botherref}
\endbibitem

%%% 56
\bibitem{search_alg_portugal_2019}
\begin{botherref}
\oauthor{\bsnm{Portugal}, \binits{R.}}:
Quantum walks and search algorithms
(2019).
\doiurl{10.1007/978-1-4614-6336-8}
\end{botherref}
\endbibitem

%%% 57
\bibitem{embedded_hypercubes_2014}
\begin{botherref}
\oauthor{\bsnm{Makmal}, \binits{A.}},
\oauthor{\bsnm{Zhu}, \binits{M.}},
\oauthor{\bsnm{Manzano}, \binits{D.}},
\oauthor{\bsnm{Tiersch}, \binits{M.}},
\oauthor{\bsnm{Briegel}, \binits{H.J.}}:
Quantum walks on embedded hypercubes.
Physical Review A
\textbf{90}(2)
(2014).
\doiurl{10.1103/physreva.90.022314}
\end{botherref}
\endbibitem


\end{thebibliography}
%% if required, the content of .bbl file can be included here once bbl is generated
%%\input sn-article.bbl

%% Default %%
%%\input sn-sample-bib.tex%

\end{document}